\documentclass{raa}            
\usepackage{natbib}
\usepackage{amssymb,amsmath}
\bibpunct{(}{)}{;}{a}{}{,}
\usepackage{graphicx,times}             
\usepackage{amsmath}
\usepackage{times}
\usepackage{graphicx}
\usepackage{graphicx}
\usepackage{epstopdf}
\usepackage{amsmath}
\usepackage{breqn}
\usepackage{color}
\usepackage{soul}
\usepackage{rotating}

\newcommand{\beq}{\begin{equation}}
\newcommand{\eeq}{\end{equation}}
\newcommand{\bea}{\begin{eqnarray}}
\newcommand{\eea}{\end{eqnarray}}

\newcommand{\gr}{$\gamma$-ray}

\begin{document}

   \title{Searching for GeV gamma-ray emission from the bulge of M31
}

   \volnopage{Vol.0 (200x) No.0, 000--000}      
   \setcounter{page}{1}          

   \author{Li Feng
      \inst{1,2}
   \and  Zhiyuan Li
      \inst{1,2}
   \and  Meng Su
      \inst{3,4}
    \and Pak-Hin T. Tam
       \inst{5}
     \and Yang Chen
        \inst{1,2}
   }

   \institute{School of Astronomy and Space Science, Nanjing University, 163 Xianlin Avenue, Nanjing 210023, China; lizy@nju.edu.cn, fengli\_sou@126.com\\
      \and 
           Key Laboratory of Modern Astronomy and Astrophysics, Nanjing University, China\\
      \and 
            Department of Physics, The University of Hong Kong, Hong Kong SAR, China\\
       \and 
            Laboratory for Space Research, The University of Hong Kong, Hong Kong SAR, China\\
       \and 
             School of Physics and Astronomy, Sun Yat-sen University, Guangzhou 510275, China\\}
   
          

\abstract{ The three major large-scale, diffuse \gr~structures of the Milky Way are the Galactic disk, a bulge-like GeV excess towards the Galactic center, and the {\it Fermi bubble}. Whether such structures can also be present in other normal galaxies remains an open question. M31, as the nearest massive normal galaxy, holds promise for spatially-resolving the \gr~emission. Based on more than 8 years of {\it Fermi}-LAT observations, we use (1) disk, (2) bulge, and (3) disk-plus-bulge templates to model the spatial distribution of the \gr~emission from M31. Among these, the disk-plus-bulge template delivers the best-fit, in which the bulge component has a TS value 25.7 and a photon-index of $2.57\pm0.17$, providing strong evidence for a centrally-concentrated $\gamma$-ray emission from M31, that is analogous to the Galactic center excess. 
The total 0.2--300 GeV $\gamma$-ray luminosity from this bulge component is $\rm(1.16 \pm 0.14)\times10^{38}{\rm~erg~s^{-1}}$, which would require $\sim1.5\times10^5$ millisecond pulsars, if they were the dominant source. We also search for a Fermi bubble-like structure in M31 using the full dataset (pass8), but no significant evidence is found. In addition, a likelihood analysis using only photons with the most accurate reconstructed direction (i.e., PSF3-only data) reveals a 4.8\,$\sigma$ point-like source located at $\sim$10 kpc to the northwest of the M31 disk, with a luminosity of $(0.97 \pm 0.27)\times 10^{38}\ {\rm erg\ s^{-1}}$ and a photon-index of $2.31\pm0.18$. Lacking of a counterpart on the southeast side of the disk, the relation between this point-like source and a bubble-like structure remains elusive.
}

   \authorrunning{L. Feng, Z. -Y. Li, M. Su, P. -H. Tam \& Y. Chen}            
   \titlerunning{Searching for GeV gamma-ray emission from the bulge of M31}  
    \maketitle

%
%
\section{Introduction}           
\label{sect:intro}

The {\it Fermi Gamma-ray Space Telescope}, with its principle instrument, the Large Area Telescope (LAT; \citealp{atwood+etal+2009}), has revolutionized our view of the $\gamma$-ray ($0.1-300$ GeV) sky since its launch in 2008.
In particular, GeV $\gamma$-ray emissions have been detected for the first time from a handful of nearby galaxies with moderate to strong star formation activities, such as the Small Magellanic Cloud (SMC; \citealp{abdo+etal+2010a}), Large Magellanic Cloud (LMC; \citealp{abdo+etal+2010b}, \citealp{ack+etal+2016}), M\,31 (\citealp{abdo+etal+2010d}, \citealp{ack+etal+2017}), M\,82, NGC\,253 (\citealp{abdo+etal+2010c}), NGC\,4945 (\citealp{ack+etal+2012a}), NGC\,1068 (\citealp{ack+etal+2012a}), NGC\,6814 (\citealp{ack+etal+2012b}), NGC\,2146 (\citealp{tang+etal+2014}) and Arp 220 (\citealp{peng+etal+2016}).
\cite{abdo+etal+2010d} found a tight correlation between the $\gamma$-ray ($\sim~0.1-100$ GeV) luminosity and the star formation rate, strongly suggesting that the GeV emission is dominated by the interaction between the cosmic-rays (CRs) and the interstellar medium (ISM).
Supernova remnants are generally thought to be the primary accelerators of CRs with energies up to $10^{15}$ eV.
{The CR hadrons can collide with the ISM to produce neutral pions}, which subsequently decay into $\gamma$-ray photons.
Leptonic processes such as inverse-Compton {and} bremsstrahlung of CR electrons may also contribute to the detected $\gamma$-ray emission (e.g., \citealp{strong+etal+2010}).

To better understand the production and transportation of CRs in galactic environments, it is desirable to spatially resolve the CR-induced, presumably diffuse $\gamma$-ray emission. However, due to the limited angular resolution of {\it Fermi}-LAT, only the nearest galaxies hold promise for such a purpose. For instance, the Magellanic Clouds have been reported to show extended GeV emission (\citealp{abdo+etal+2010a}, \citealp{abdo+etal+2010b}, \citealp{ack+etal+2016}).
Located at a distance of 780 kpc \citep{stan+etal+1998}, the Andromeda galaxy (M31) is perhaps the only massive external galaxy that currently permits a spatially-resolved study with the {\it Fermi}-LAT. Indeed, with an inclination angle of $\sim$78$^\circ$, the HI disk of M31 spans 3.2\degr$\times$1\degr\ on the sky, which makes M31 a potentially resolvable source to the Fermi-LAT (LAT's single-photon resolution, FWHM $\approx$ 0.8$^\circ$, for a $\geqslant$1~GeV photon).

{The} $\gamma$-ray emission from M31 has been the focus of various recent works (e.g. \citealp{abdo+etal+2010d}, \citealp{li+etal+2016}, \citealp{psh+etal+2016}). Using the {first two years LAT pass6 data, \cite{abdo+etal+2010d} first detected the GeV emission from M31, which is spatially correlated with the IRAS 100~$\mu$m image}, a good tracer of the neutral gas primarily located in the disk of M31.
\cite{psh+etal+2016}, {who used 7-year LAT data, claimed a detection of halo structures similar to the Fermi bubbles in our Galaxy \citep{su+etal+2010}.} In particular, \cite{psh+etal+2016} adopted a template of two 0.45$\degr$-radius uniform circular disks, {which are} symmetrically located perpendicular to the M31 disk, and derived a total $0.3-100$ GeV luminosity of (3.2$\pm$0.6)$\times 10^{38}{\rm~erg~s^{-1}}$ {from these two disks}. \cite{bird+etal+2015} studied the $\gamma$-ray emission from M31 using VERITAS observations and {6.5-year pass7 data of {\it Fermi}-LAT.} {Their} 54-hour VERITAS observations {had resulted} in an upper limit of the $\gamma$-ray flux above 100 GeV, while {their {\it Fermi}-LAT spectrum suggested} a turnover below $\sim$1 GeV.

More recently, \cite{ack+etal+2017} performed a detailed morphological analysis using $1-100$ GeV photons detected by {\it Fermi}-LAT in the first 7 years. They tested different morphological representations of M31: a central point source, a Herschel map, a Spitzer map, a neutral hydrogen column density map, a projected uniform circular disk on the sky, a projected elliptical disk, a Gaussian disk, and an elliptical Gaussian disk. As the authors admit, it remains inconclusive {to give the best spatial template statistically. For simplicity, a uniform circular disk (as projected on the sky)  was adopted as the spatial model of M31 by \cite{ack+etal+2017}.} They also concluded that the GeV emission of M31 might be more confined to the inner regions than a uniform circular disk template would predict. They suggested that the emission is not correlated with regions rich in gas or star formation activity, and gave an alternative and non-exclusive interpretation that the emission results from a population of millisecond pulsars(MSPs) dispersed in the bulge and disk of M31 by disrupted globular clusters or {from the decay/annihilation} of dark matter particles, as an analogy to what have been proposed to account for the Galactic center excess {found by {\it Fermi}-LAT.}

In the Milky Way, GeV excess in the Galactic center has been extensively examined  (\citealp{zhou+etal+2015}, \citealp{cal+etal+2015}, \citealp{aj+etal+2016}, \citealp{dsy+etal+2016}). There are mainly two explanations for this excess: the dark matter (DM) annihilation scenario and the astrophysical scenario, {the latter one usually involving unresolved MSPs.} \cite{hooper+etal+2013} have performed a series of work on the DM annihilation origin of Galactic center $\gamma$-ray excess. They argued that the millisecond pulsar scenario cannot explain all the excess emission, and the argument seems to have received support {from a detailed study of the MSPs in Galactic globular clusters }\citep{hooper+etal+2016}. 
However, the alternative, astrophysical scenario has gained more support from various groups over the recent years: the scenario including the Galactic center CRs (\citealp{cholis+etal+2014}), the MSPs in the bulge (\citealp{yuan+etal+2014}), and the disrupted globular clusters (\citealp{brandt+etal+2015}). \citet{yang+etal+2016} noted that the GeV excess in Galactic center shows no spherical symmetry, but rather a bipolar distribution, which may indicate an astrophysical origin. Most recently, \citet{2018NatAs...2..387M} explained the excess with the X-shaped stellar bulge and the nuclear bulge in the Galactic center, and strongly preferred an astrophysical origin rather than a DM origin.

{Since the $\gamma$-ray photons are optically thin in the {\it Fermi}'s eye, the Galactic plane is a projection of 3D distribution of $\gamma$-ray emission.} It would reveal more information if one could measure this excess from a location outside of the Milky Way.  A GeV excess from the center of M31 (analogous to the excess towards the Galactic Center), if present could also be potentially detected by {\it Fermi}. {\it In this work, we specifically search for such an excess towards the M31 center, which we refer to as a bulge component.}  We utilize more than 8-year {\it Fermi}-LAT data to provide further insight on the origin of GeV emission from M31. Our data reduction procedure and analysis are presented in Sect. 2. We discuss the possible origins of the $\gamma$-ray emission from M31 in Sect. 3, and summarize our study in Sect. 4. 
\section{Observations and data analysis} 
\label{sec:jobs}

\subsection{Data preparation} 
\label{subsec:data}

\begin{figure}
\includegraphics[width=0.5\textwidth]{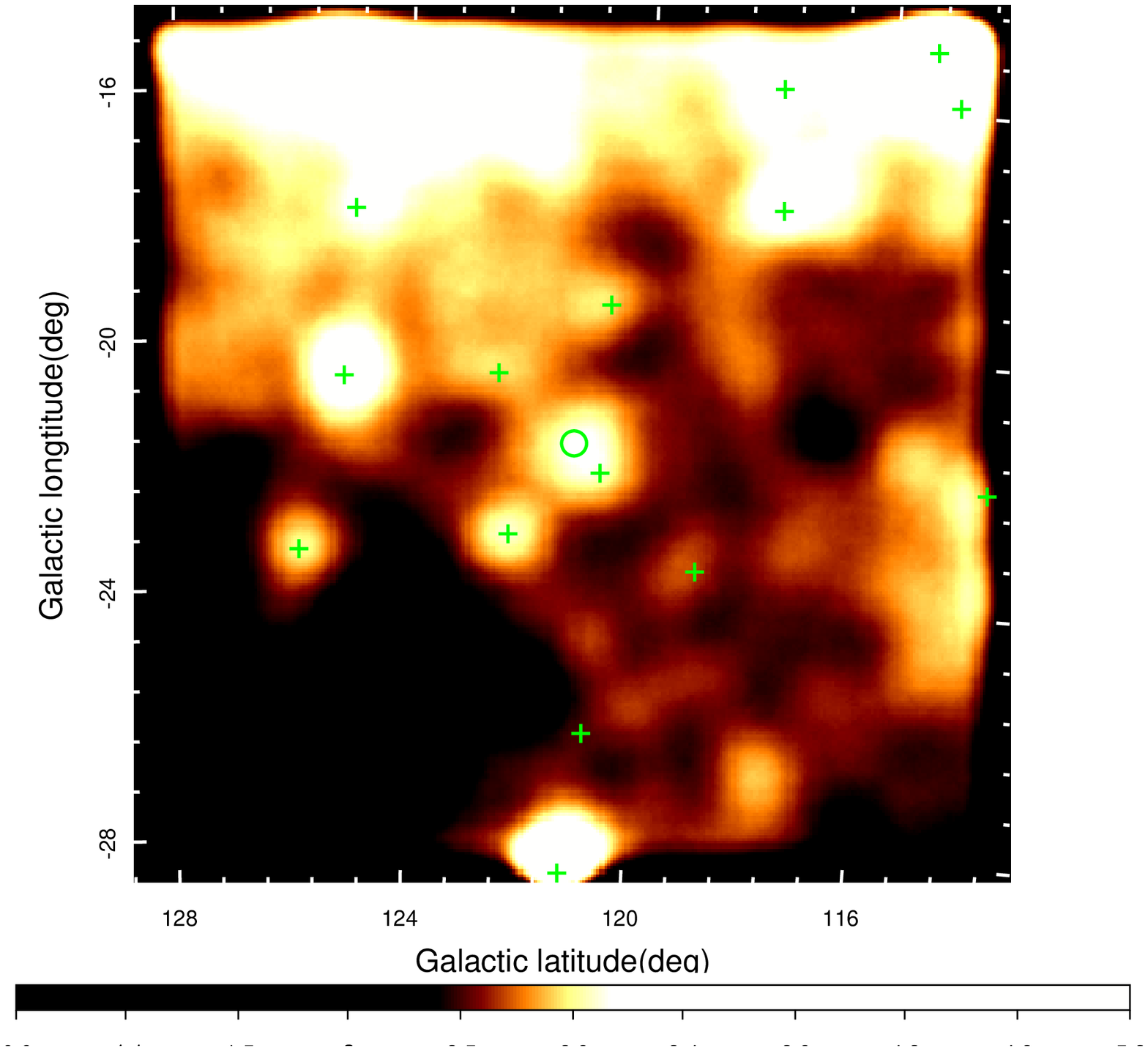}
\includegraphics[width=0.5\textwidth]{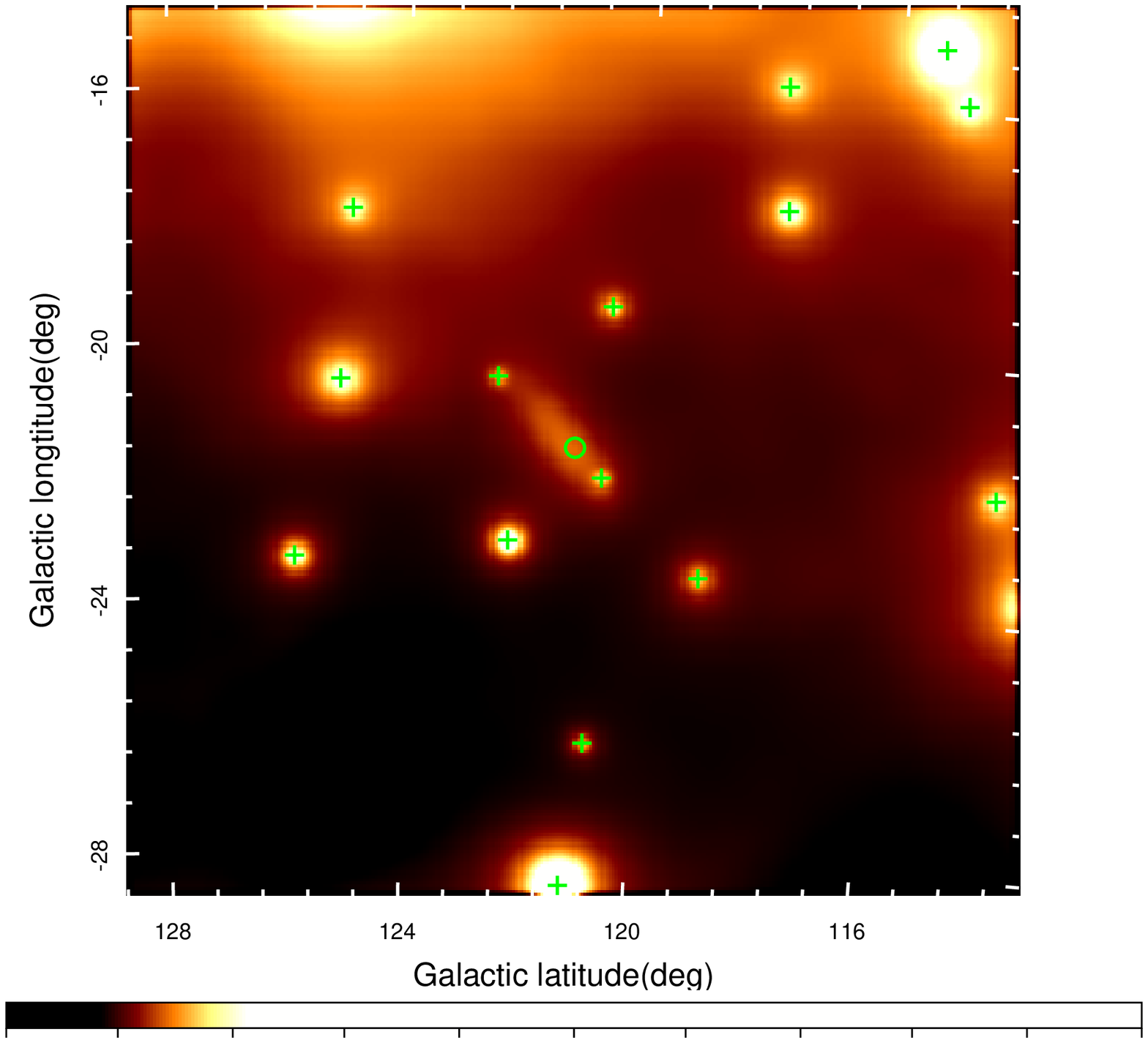}
\caption{{\it Left:} 0.2--300 GeV counts map of the M31 field, ROI = 14\degr$\times$14\degr, smoothed with a Gaussian kernel of 0.8\degr. {\it Right:} {Counts map of the background model of the ROI. In both panels, the} background point sources (extracted from 3FGL catalogue) are marked as green crosses. The center of M31 is marked as a green circle.}
\label{fig:ROI}
\end{figure}

Our analysis is based on the data taken by the {\it Fermi}-LAT between August 8, 2008 and October 7, 2016, over a period of more than 8 years.
{The {\it Fermi} Science Tools v10r0p5 is used in our analysis, and the data used here are restricted to the ones with zenith angles $<$ 100$\degr$, and within the time intervals when the satellite rocking angle was less than 52$\degr$.}
{We include all the $0.2-300$ GeV events within a rectangular region of interest (ROI), with a size of 14$\degr$~$\times$ 14$\degr$~centered at M31 [RA, DEC] = [00h42m44.3s, 41\degr16\arcmin09\arcsec] (see Figure~\ref{fig:ROI}). Our background model} includes the 3FGL catalog sources (95 sources within a radius of 20$\degr$ from the center of M31), the Galactic diffuse emission ($ \it \rm gll\_iem\_v06.fits$),  and the isotropic emission ($\it \rm iso\_P8R2\_SOURCE\_V6\_v06.txt$), using the user-contributed {\it \rm make3FGLxml.py} tool. The adopted instrument response function (IRF) is $\it \rm P8R2\_SOURCE\_V6$. 

\subsection{Analysis} \label{subsec:anal}
\subsubsection{Spatial models of M31} \label{subsec:spm}
As we are most interested in distinguishing the physical regions from where the observed $\gamma$-ray emission is produced, {we use three spatial templates to model the $\gamma$-ray emitting region of M31: the disk only, the bulge only and the disk+bulge templates. Assuming a hadronic origin, we employ the IRAS 100~$\mu$m image as the spatial model for the disk component in both the disk and disk+bulge templates, where IRAS 100~$\mu$m is a good tracer of the neutral gas.
We use a power law as the spectral model for the disk component. During the analysis, we set free the spectral parameters of M31 and any 3FGL sources with a distance to M31 of $<$10$\degr$, we also set free the normalization parameters of Galactic diffuse emission and isotropic background emission.} Positions of the background sources are fixed to those given in the 3FGL catalog. 

A point source was employed to model the bulge component {in both the bulge and the bulge+disk templates.} In principle, the bulge component should be extended, if the $\gamma$-ray emission predominantly arises from stellar populations and/or dark matter.
{Due to the small spatial extension of the M31 bulge, i.e., a high-light radius of $\sim$1 kpc, or $\sim$4$\arcmin$ is suggested by \citep{dong+etal+2015}, a point source spatial model could perform well to represent the $\gamma$-ray emission from the bulge by LAT.}
We note that the central super-massive black hole in M31 is currently extremely quiescent (Li et al.~2011), thus no significant $\gamma$-ray emission from an AGN is expected. 

\begin{figure}[tbh!]
\centerline{
\includegraphics[width=0.5\textwidth]{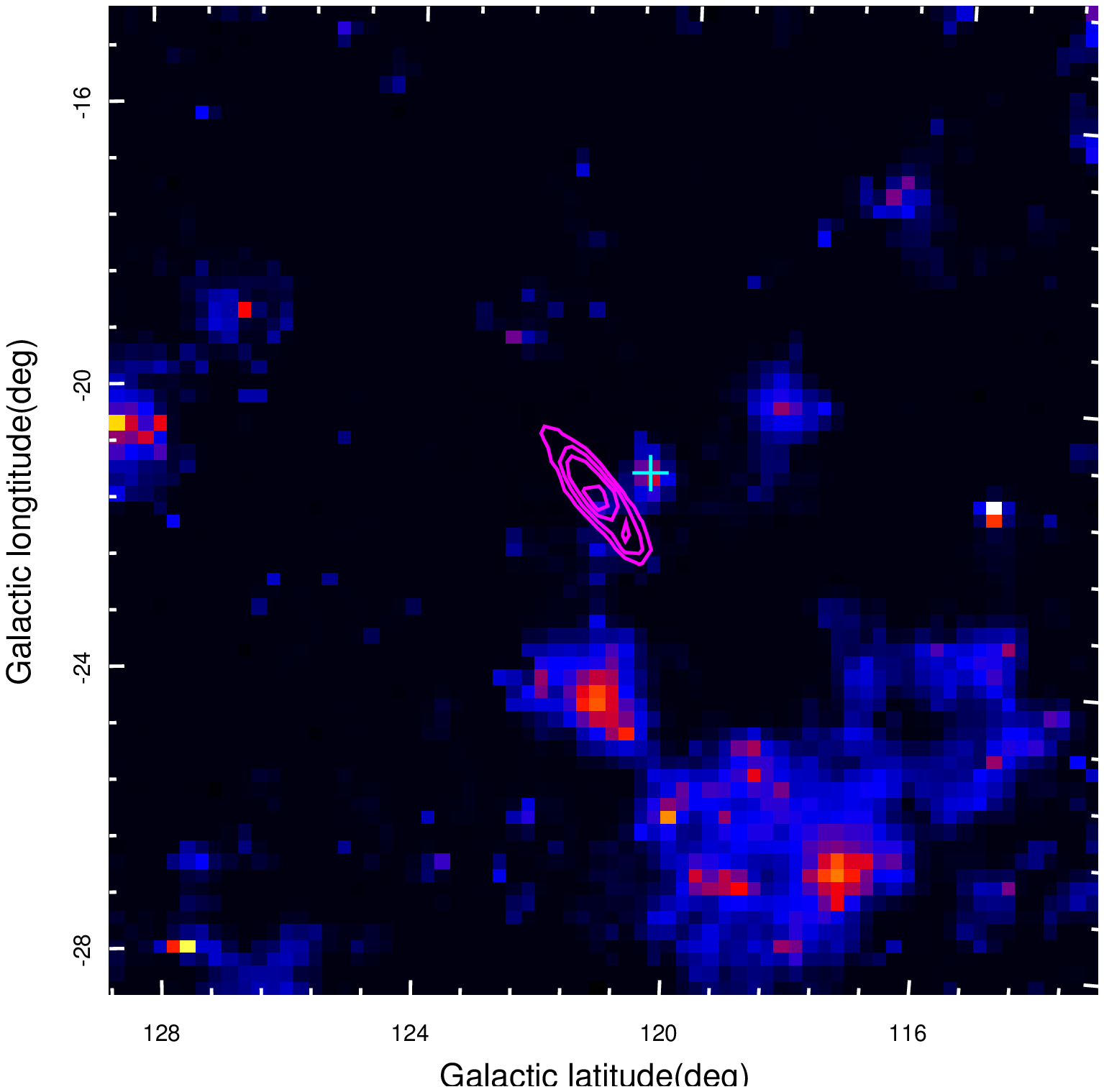}
\includegraphics[width=0.5\textwidth]{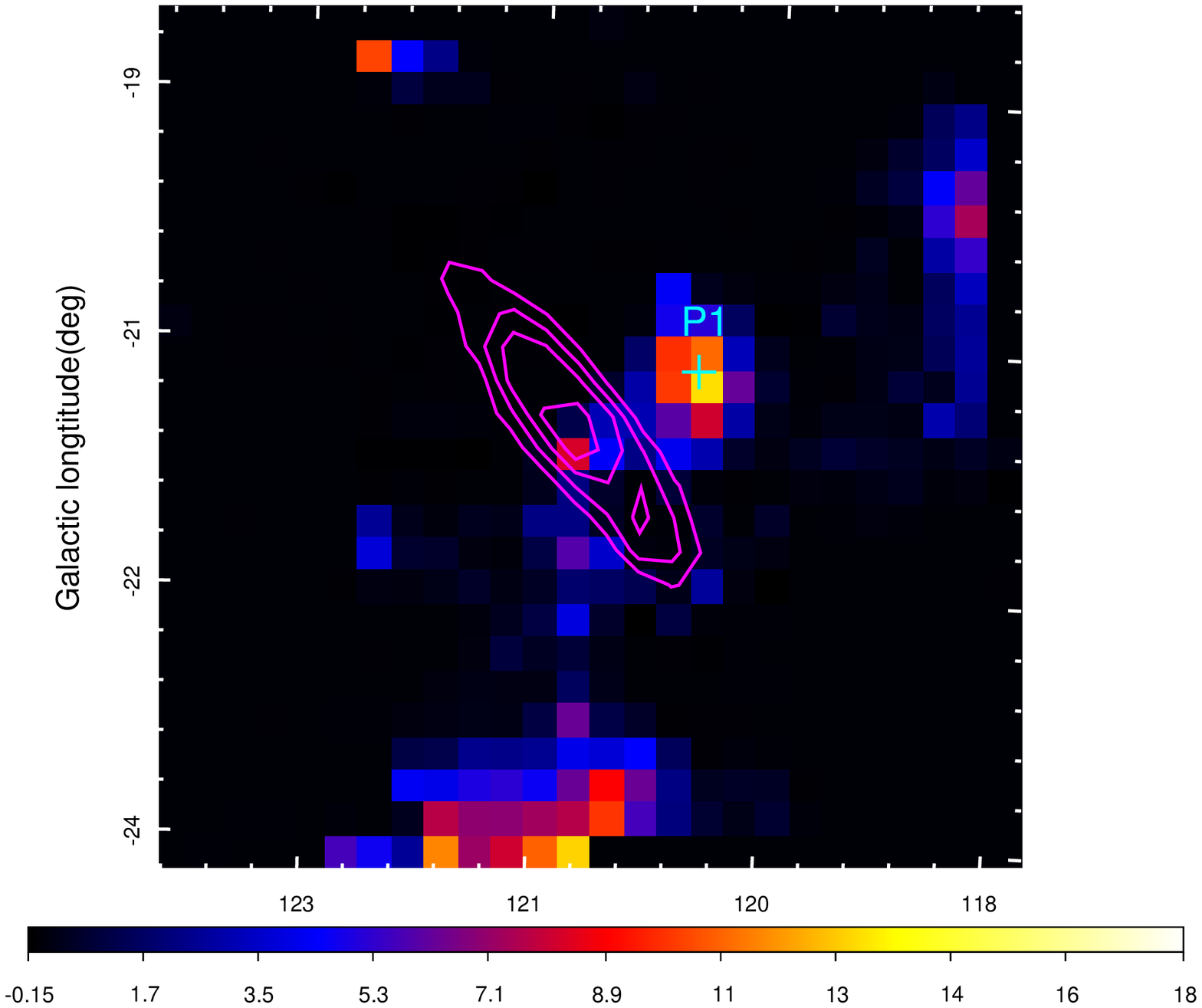}
}
\vskip0.7mm
\centerline{
\includegraphics[width=0.5\textwidth]{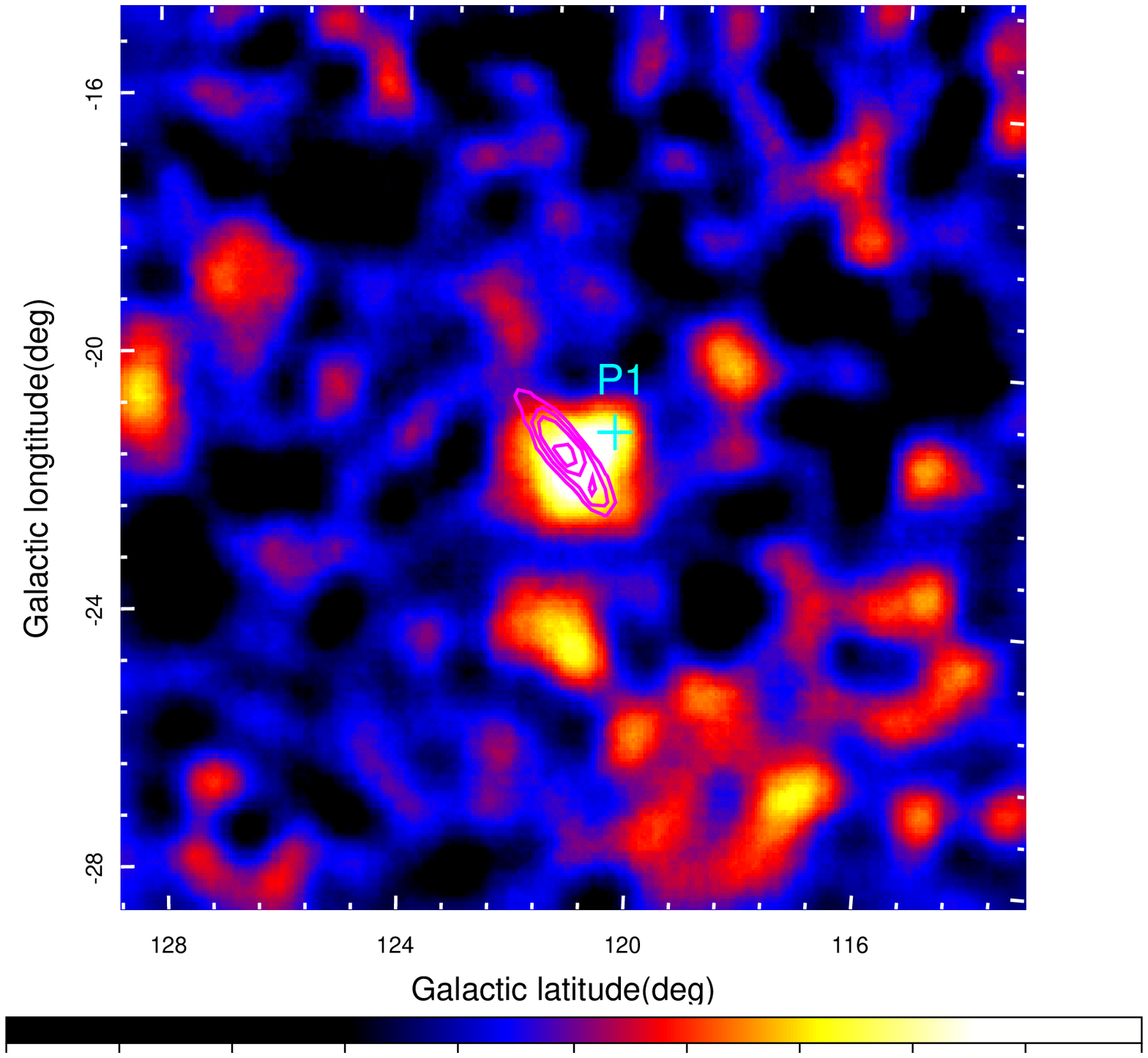}
\includegraphics[width=0.5\textwidth]{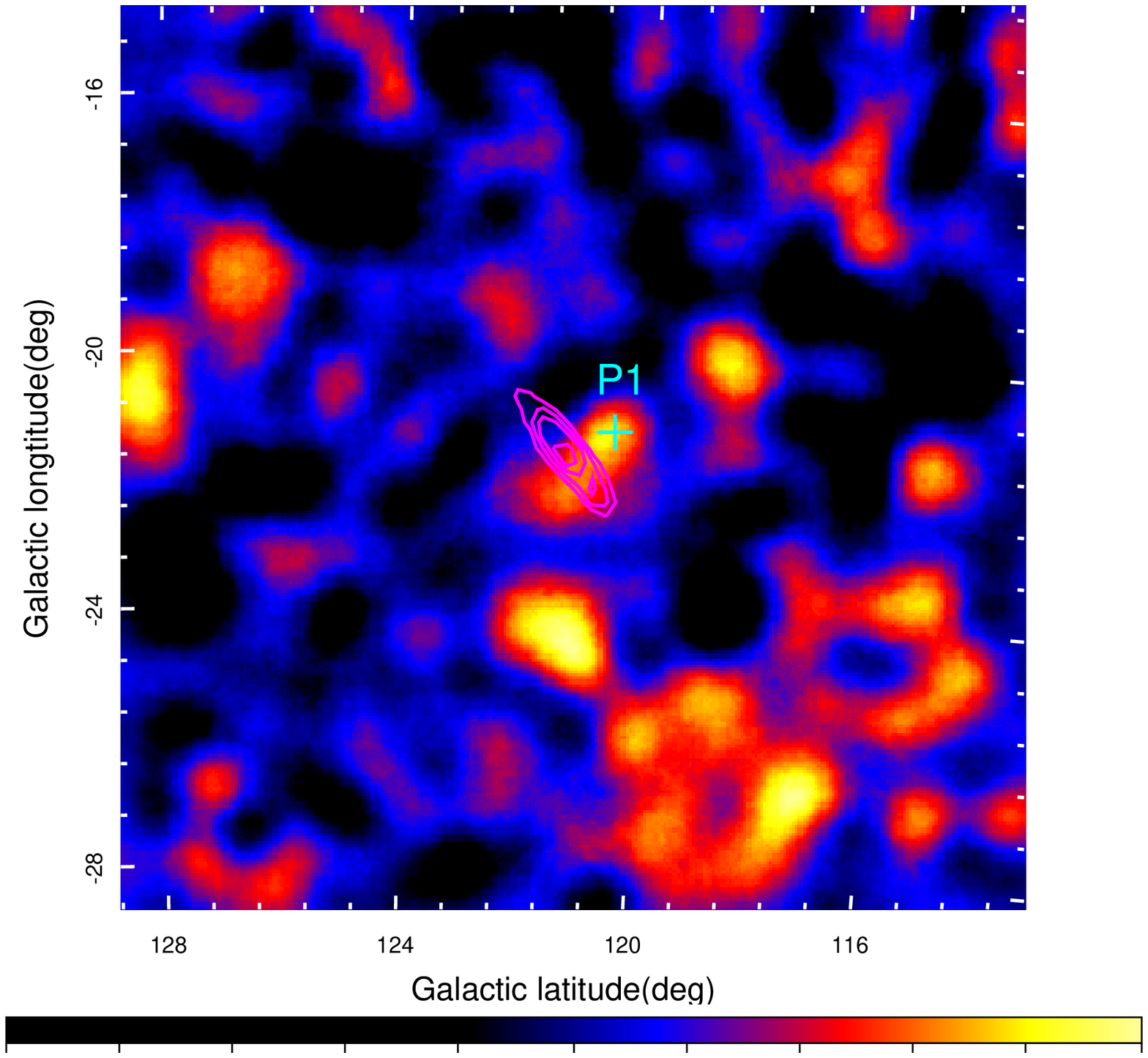}
}
\caption{TS maps ({\it top panels}) and residual counts maps ({\it bottom panels}) with the pass8 PSF3 data. The disk-only model is adopted, and P1 is taken as a point source. In each panel, all background sources have been subtracted; the IRAS 100~$\mu$m intensity is shown with magenta contours, and P1 is marked as a cyan cross. {\it Top left:} TS map of ROI = 14\degr$\times$14\degr, without subtracting the disk model; {\it Top right:} Zoom-in of the TS map of the 5$\degr\times5\degr$ rectangular region, with the disk model further subtracted. The yellow cross marks the optical center of M31, and the green/white circle represents the 1-$\sigma$ error circle of the best-fit centroid position of M31, assuming the bulge model using 0.2--300 GeV/1--300 GeV data, respectively (Section~\ref{subsec:spm}). The blue diamond marks the position of NGC\,205, while the green cross marks the position of $\rm FL8Y J0039.8+4204$. {\it Bottom left:} The residual counts map without the disk model subtracted; {\it Bottom right:} The residual counts map with the disk  model subtracted.}
\label{fig:psf3}
\end{figure}

We use power law as the spectral model for the bulge template. {As seen in Table~\ref{tab:test}, at an energy band of 0.2--300 GeV, the best fit position of M31 is [RA, DEC] = [10.7806$\degr$, 41.2741$\degr$] with error radius$\sim$0.09$\degr$, and the optical center of M31 of [RA, DEC] = [10.6847$\degr$, 41.2687$\degr$] is well within such an error circle, see Figure~\ref{fig:psf3} (top right, the green circle).} We notice that \cite{ack+etal+2017} did the same test in an energy band of $>$ 1GeV. Using their energy selection, we found a best fit position of M31 as [RA, DEC] = [10.8466$\degr$, 41.2223$\degr$], with a radius of error circle $\sim$ 0.0769$\degr$, and the optical center of M31 is slightly outside this error circle~(see Figure~\ref{fig:psf3}, top right, the white circle), which agree with \cite{ack+etal+2017}. In the following analysis, we fix the point source to be the center of M31.

\begin{figure}[tbh!]
\centerline{
\includegraphics[width=0.5\textwidth]{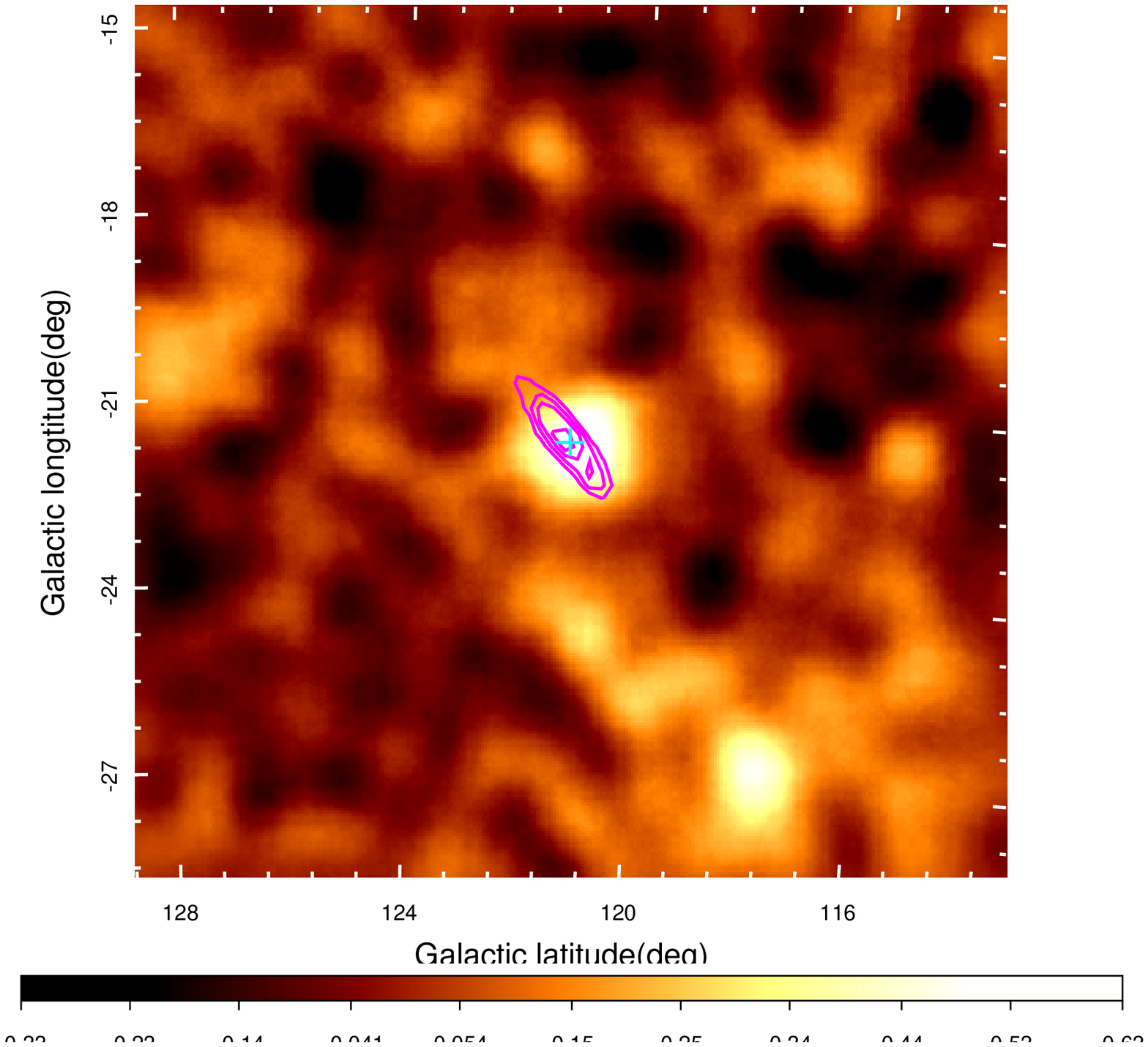}
\includegraphics[width=0.5\textwidth]{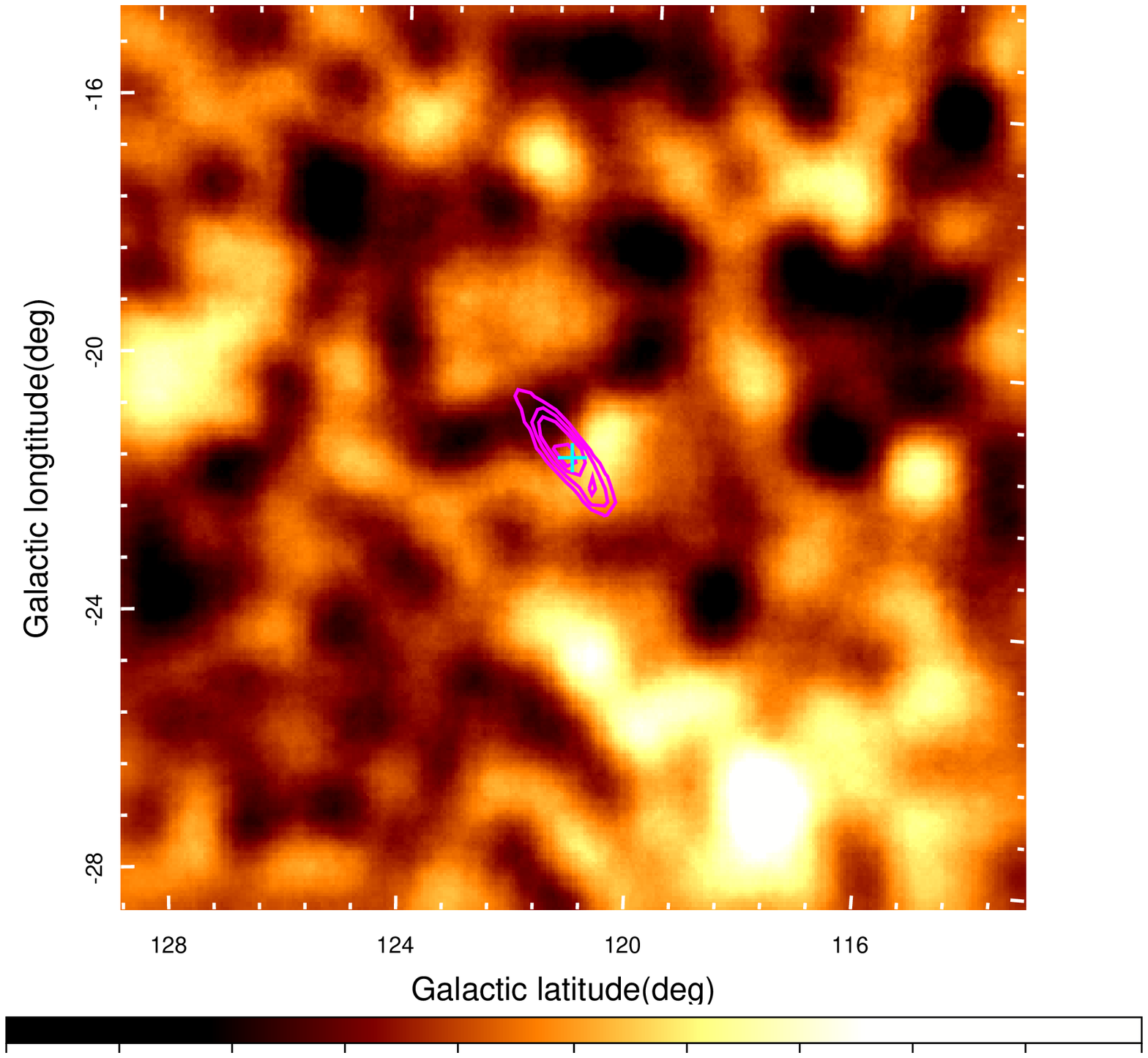}
}
\vskip0.7mm
\centerline{
\includegraphics[width=0.5\textwidth]{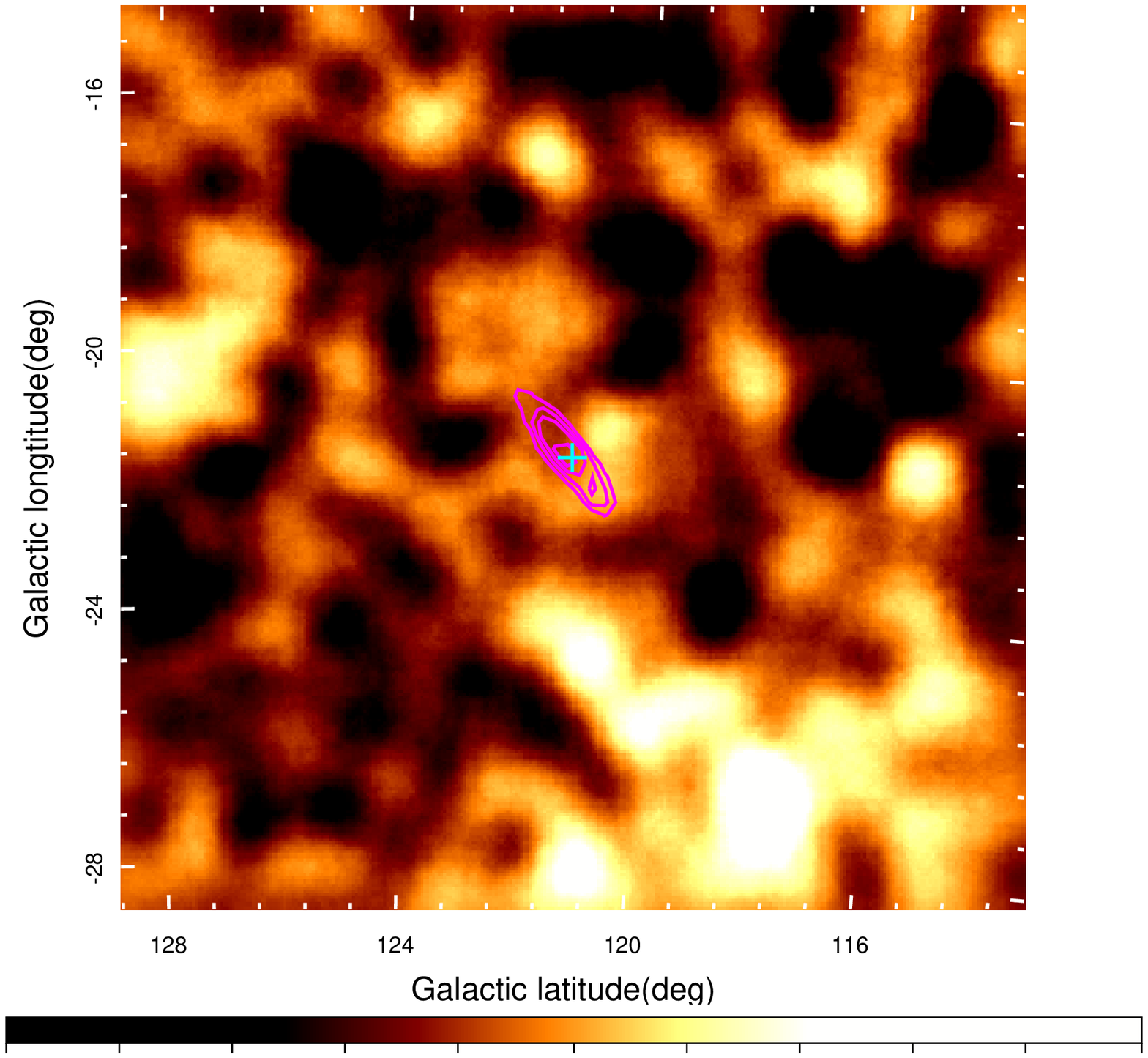}
\includegraphics[width=0.5\textwidth]{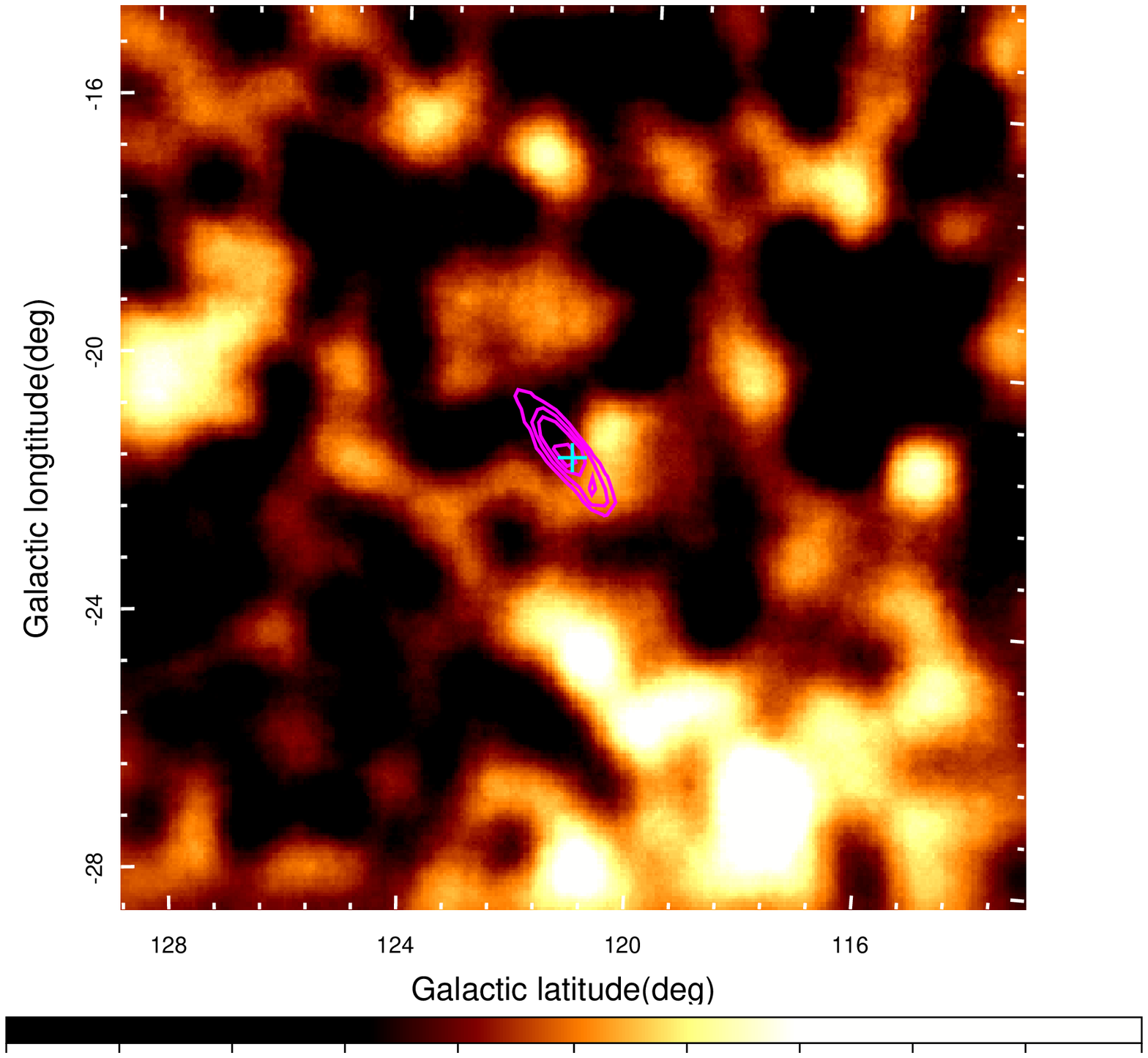}
}
\caption{0.2--300 GeV residual counts maps. In all panels, the IRAS 100\,$\mu$m intensity contours (magenta) are overlaid, and the cyan cross marks the center of M31. All maps are smoothed with a Gaussian kernel of 0.8\degr. {\it Top left:} only background sources are subtracted. Prominent emission is clearly seen coincident with M31.
{\it Top right:} the disk model and background sources are subtracted. {\it Bottom left:} the bulge model and background sources are subtracted. {\it Bottom right:} the disk+bulge model and background sources are subtracted. 
}
\label{fig:models}
\end{figure}

Figure~\ref{fig:models}~shows the $0.2-300$ GeV background-subtracted counts maps derived with the different spatial models of M31, and all maps are overlaid with IRAS 100\,$\mu$m intensity contours. Emission from M31 is clearly visible (top left panel in Figure~\ref{fig:models}). At a glance, all three spatial models lead to a reasonable characterization of M31 (other panels in Figure~\ref{fig:models}).

{To further evaluate the goodness of different templates, we examine the log(likelihood) (denoted by log$~\mathcal{L}$~in the following) of each fitting, to find the maximum likelihood goodness-of-fit.} We note that the background model, i.e., excluding any components of M31, gives log$~\mathcal{L}_{\rm B}$ = -501420.
Taking this as the fiducial value, an increase in log$~\mathcal{L}$~when one adds a source model component (i.e., M31) indicates a more significant improvement of the fit (Ackermann et al. 2017).
{As seen in table~\ref{tab:pass8},} log$~\mathcal{L}$ of the disk, bulge and disk+bulge spatial template is -501389, -501400 and -501383, respectively. This suggests that the disk+bulge template is more favored with a significance $>3\sigma$. All three templates predict similar 0.2-300 GeV luminosities. The significance of each spatial component is expressed by a test statistic (TS) value, $\rm TS = 2(log~\mathcal{L}-log~\mathcal{L}_{\rm B})$. {In the fitting results of the disk+bulge template},  $\it TS_{\rm disk} = 33.4$, $\it TS_{\rm bulge} = 25.7$, which strongly suggests the detection of the bulge emission from M31. 

To compare with the uniform disk template used by \cite{ack+etal+2017}, we also test the uniform disk template. The difference of log(likelihood) between the best-fitted uniform disk template (radius $\sim0.5\degr$) and the disk+bulge template is not significant ($<2\sigma$) for 1--300 GeV data.  For comparison, \cite{ack+etal+2017} declared a uniform disk with radius $\sim0.38\degr$ best fitted the data. We also test the uniform disk model  with 0.2--300 GeV data, which has a best-fit radius of $\sim0.9\degr$ and a similar significance. However this might be due to the energy-dependent PSF. We note that compared the uniform disk model,  the disk+bulge template is more physically motivated.

\begin{table*}
\caption{Likelihood analysis for pass8 data (0.2--300 GeV)}
\begin{tabular}{ccccccc}
  \hline
  \hline
Spatial model & Composition & $\Gamma$ & $\rm F_{0.2-300~GeV}$ & $\rm L_{0.2-300~GeV}$ & TS value & log$~\mathcal{L}$  \\

  &  &  & ($10^{-9}{\rm~ph~s^{-1}~cm^{-2}}$) & ($10^{38}{\rm~erg~s^{-1}}$) &   &    \\
(1)& (2) & (3) & (4) & (5) & (6) & (7) \\
\hline\hline

disk    & IRAS 100$\mu$m   & 2.31$\pm$0.09    & 4.60$\pm$0.56 &    4.0$\pm$0.52   &  107.87&  -501389\\
 \hline
bulge &        &2.55$\pm$0.11      & 3.01$\pm$0.40  &   2.16$\pm$0.29  & 82.90 &    -501400\\
\hline
disk+bulge &IRAS 100$\mu$m       & 2.22$\pm$ 0.13   & 2.45$\pm$0.24  &   2.48$\pm$0.32    & 33.40      & -501383\\
           & bulge     & 2.57$\pm$ 0.17     & 1.83$\pm$0.26   &    1.16$\pm$0.14    &  25.70        &        -      \\
\hline
 disk + $\rm P1 $ & IRAS 100$\mu$m     &2.26$\pm$0.09    &3.86$\pm$0.59  &    $3.65\pm$0.55   &83.63    &-501378\\
              &P1           &2.32$\pm$0.18     &1.12$\pm$0.41  &   0.97$\pm$0.27    & 23.23&   -      \\
\hline
 bulge+ $\rm P1 $& bulge &2.51$\pm$0.12    &2.90$\pm$0.54  &    $1.94\pm$0.30   &  66.77&   -501379\\
                   &P1  &2.30$\pm$0.20     &1.09$\pm$0.45  &   0.97$\pm$0.20    & 23.20    &  -      \\
 \hline
 disk + bulge+  P1  & IRAS 100$\mu$m           &2.18$\pm$0.13       & 2.12$\pm$0.52   &   2.33$\pm$0.56    &     27.86      & -501373  \\
                    & bulge    &  2.51$\pm$0.18   &  1.53$\pm$0.46    &   1.01$\pm$0.27  &  17.68     &  -         \\
                    & P1   &2.29$\pm$0.19    & 1.04$\pm$0.40  &    0.94$\pm$0.25    & 22.06  &    -           \\
\hline

\label{tab:pass8}
\end{tabular}
\end{table*}

\begin{table*}
\caption{Likelihood analysis for pass8 PSF3 data (0.2--300 GeV)}
\begin{tabular}{ccccccccc}
\hline
\hline
spatial model  & composition & $\Gamma$ & $\rm F_{0.2-300~GeV}$  & $\rm L_{0.2-300~GeV}$  & TS value & log$~\mathcal{L}$ \\
& &
&
($10^{-9}{\rm~ph~s^{-1}~cm^{-2}}$) &
($10^{38}{\rm~erg~s^{-1}}$) &
 &
\\
(1) &
(2) &
(3) &
(4) &
(5) &
(6) &
(7) \\
\hline
\hline
disk      & IRAS 100$\mu$m    & 2.36$\pm$0.14    & 4.78$\pm$0.75   &  3.88$\pm$0.78   &  62.48   &  -213378 \\
 \hline
bulge  &   &2.71$\pm$0.17      & 3.61$\pm$0.63 &    2.01$\pm$0.34  &   55.02   &   -213374 \\
\hline
disk+bulge & IRAS 100$\mu$m     & 2.06$\pm$ 0.3     & 1.6$\pm$ 0.69  &    2.30$\pm$0.40   &  26.76      &   -213371 \\
             & bulge & 2.76$\pm$ 0.22     & 2.69$\pm$ 0.24   &   1.42$\pm$0.1   &  10.38    &   -           \\
\hline
disk+P1    & IRAS 100$\mu$m      & 2.26$\pm$ 0.15   &   3.89$\pm$ 0.80  &    3.66$\pm$0.85   & 46.45   &   -213366 \\
             & P1    & 2.52$\pm$ 0.26   & 1.66$\pm$ 0.63   &   1.09$\pm$0.33   &  19.60       &  -       \\
\hline
bulge+ P1 & bulge     &   2.64$\pm$ 0.18   & 2.99$\pm$0.60   &    1.75$\pm$0.34   &  41.92 &   -213366 \\
          & P1      &  2.46$\pm$ 0.26    &  1.48$\pm$ 0.56   &   1.04$\pm$0.32   &  17.77   &  -            \\
\hline
disk + bulge+  P1 & IRAS 100$\mu$m       &2.01$\pm$0.20  & 1.44$\pm$0.60       & 2.33$\pm$1.10 &  9.65    & -213363  \\
                  & bulge      &  2.68$\pm$0.21  &  2.98$\pm$0.70    & 1.75$\pm$0.36 &  19.12&  -            \\
                  & P1    &2.47$\pm$0.22     & 1.47$\pm$0.61  &    1.04$\pm$0.34    &  17.86 &   -         \\

\hline
\label{tab:psf3}
\end{tabular}
\end{table*}

\subsubsection{Energy dependent analysis} \label{subsec:data}
We further divide the {full data (pass8) by} 3 energy bands: $0.2-1$, $1-20$ and $20-300$ GeV. With each sub-band data, we perform the likelihood analysis to study the $\gamma$-ray morphological distribution. The background-subtracted counts maps in the sub-bands are shown in Figure~\ref{fig:subband}. On the $0.2-1$ GeV and $1-20$  GeV counts maps, there is significant emission {from} the center of M31.

We employ the three spatial models in the morphological fitting of each sub-band as well. The parameters of each model are listed in Table~\ref{tab:subband}. In the $0.2-1$ GeV sub-band, log$~\mathcal{L}$ values of the disk, bulge, disk+bulge model are very close to each other{, although} the disk+bulge model is slightly better than the disk or bulge model. In the $1-20$ GeV band, {the disk+bulge model has shown significantly better result than the other models do.} In the $20-300$ GeV sub-band, the TS value of M31 is effectively zero in all three models, implying that the $\gamma$-ray emission from M31 is insignificant above 20 GeV. 

\subsubsection{PSF3 analysis} \label{subsec:psf3} 
Angular resolution of data is the key to morphological studies. In pass8 data, PSF type (PSF0, PSF1, PSF2, and PSF3) refers to the quality of reconstruction of direction of photons, with PSF3 having the best accuracy. To search for possible substructures of M31 under an improved angular resolution, we carry out likelihood analysis {using only type PSF3 data.}
We select the 0.2--300 GeV data between August 8, 2008 and October 7, 2016, which is the same as before. The data are restricted to the ones with zenith angles $<$ 100$\degr$, and within the time intervals when the satellite rocking angle was less than 52$\degr$. We also restrict the data to a rectangular region of interest (ROI), with a size of 14$\degr$~$\times$ 14$\degr$~centered at M31. Our background model includes the 3FGL catalog sources (95 sources within a radius of 20$\degr$ from the center of M31), the Galactic diffuse emission ($ \it \rm gll\_iem\_v06.fits$),  and the isotropic emission ($\rm iso\_P8R2\_SOURCE\_V6\_PSF3\_v06.txt$ ). The adopted instrument response function (IRF) is $\it \rm P8R2\_SOURCE\_V6::PSF3$. The results using only PSF3 data are presented in Table~\ref{tab:psf3}. It can be seen that the log$~\mathcal{L}$ of disk+bulge model is {higher than the ones of bulge model and disk model}, which is consistent with the ones using the full data set. It is not surprising that the TS values of the spatial models {using only PSF3 data} are less than the ones using the full data (Table~\ref{tab:pass8}), as the number of photons is smaller.

In the residual map generated from the analysis of pass8 PSF3 data (Figure~\ref{fig:psf3}), we can identify a point-like source at [RA, DEC] = [00h39m12s, 41$\degr$39$\arcmin$36$\arcsec$], {with a distance of $\sim$ 0.7$\degr$ to the center of M31} in the northwest. Hereafter we designate this source candidate as P1. \cite{li+etal+2016} find a point-like {GeV} excess at (00h39m48s, 41$\degr$52$\arcmin$00$\arcsec$), and Ackermann et al. (2017) find `excess2' (00h40m00s, 42$\degr$07$\arcmin$48$\arcsec$); both their locations are roughly in accordance with the location of P1. In addition, there is a nearby point source  FL8Y J0039.8+4204 in FL8Y catalogue\footnote{$\rm https://fermi.gsfc.nasa.gov/ssc/data/access/lat/fl8y/gll\_psc\_8year\_v5\_assoc.reg$}, however {it is not spatially coincident with P1} (Figure~\ref{fig:psf3}).

Therefore, we add P1 to the source model file and redo the likelihood analysis, with {PSF3 data and new} spatial models: disk+P1, bulge+P1, disk+bulge+P1. The corresponding results are shown in the last three rows of Tables~\ref{tab:psf3}. The disk+bulge+P1 model has the {highest log$~\mathcal{L}$. Adding P1 has improved the log$~\mathcal{L}$ for all three previous spatial models.} We also redo {the likelihood analysis on the three new spatial models} with pass8 full data set, see Section \ref{subsec:bubble}, as a higher photon statistics may improve the detection significance of P1.

We noticed that NGC\,205 (M110), a satellite dwarf galaxy of M31, [RA, DEC] = [00h40m22.1s, 41$\degr$41\arcmin07\arcsec], lies close to the position of P1 (offset by $\sim$0.3 $\degr$, see Figure~\ref{fig:psf3} top right). To see if P1 could be the counterpart of {this} dwarf galaxy NGC 205, we {redo the} likelihood analysis, with NGC\,205 added as a new point source. Firstly, we replace P1 with NGC\,205, this reports a TS value of 12.1 for NGC 205, which is much less than the TS value of P1 in previous models. Then we put both P1 and NGC\,205 into the model. In this case we have $\rm TS_{P1} = 21.67$ and $\rm TS_{NGC\,205} = 0.06$. The results suggest that NGC\,205 {may not be} responsible for the excess emission.

\subsection{Testing the existence of bubble-like feature of M31}

\cite{psh+etal+2016} had performed a search for extended $\gamma$-ray halo around M31. They reported a 5.2$\sigma$ significance for two 0.45$\degr$ circles model, and a 4.7$\sigma$ significance for 0.9 circle model.  \cite{li+etal+2016} also applied 0.9$\degr$ circle and two 0.45$\degr$ circles as spatial model of excess emission around M31, as well as point source model. But they did not find {any} significant bubble-like features. 

\begin{figure}
\includegraphics[width=0.6\textwidth]{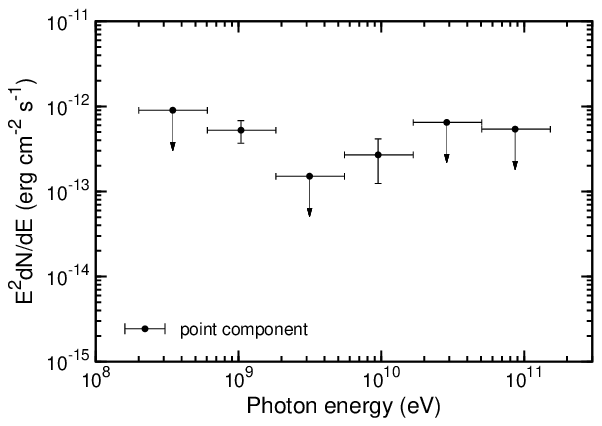}
\includegraphics[width=0.4\textwidth]{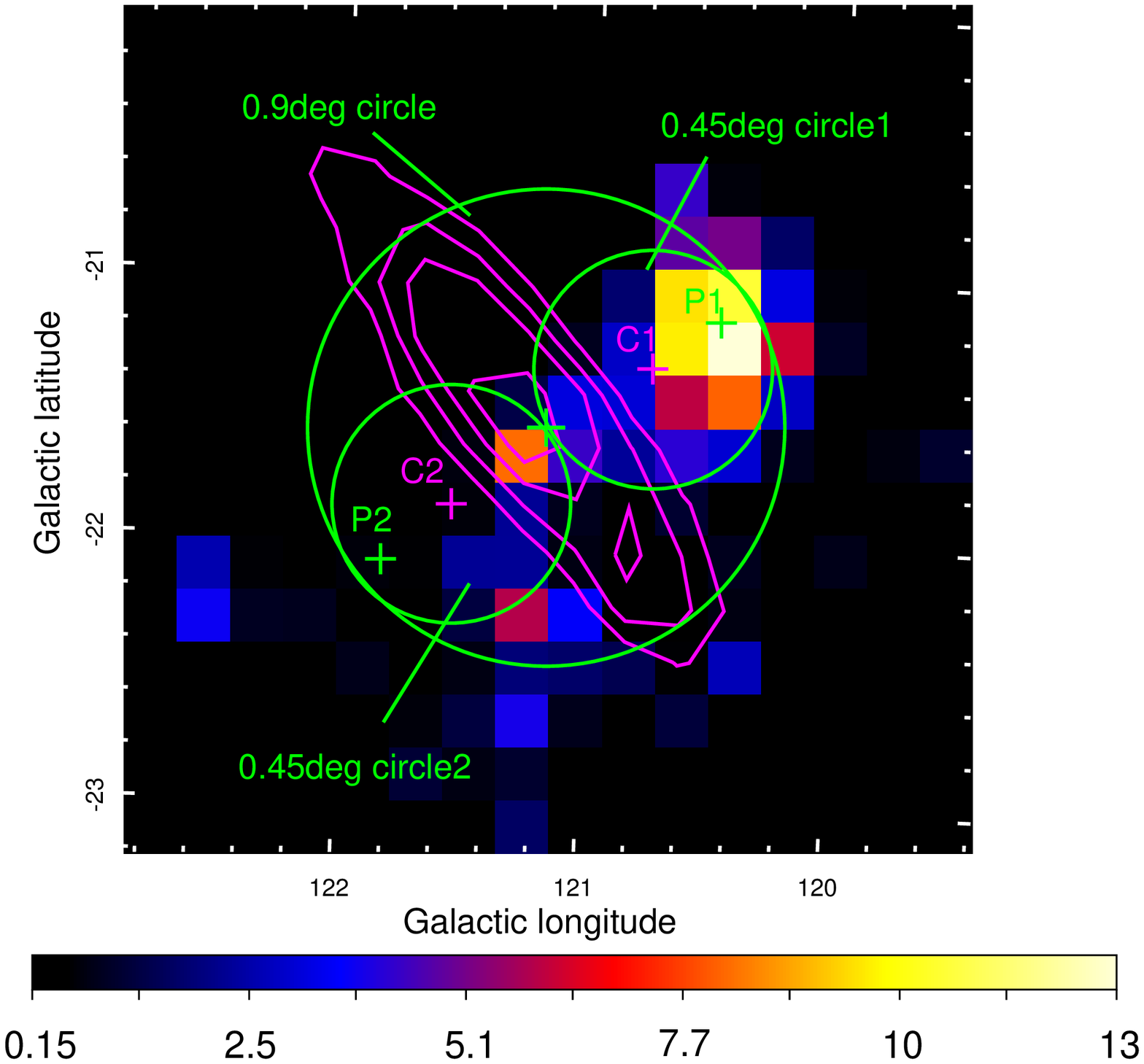}
\caption{ Fitting results of P1 with pass8 data and the disk model. {\it Left panel:} SED of P1. {\it Right panel:} $3\degr\times3\degr$~TS map generated with a source model including all background sources and the disk model. Overlaid are the IRAS 100\,$\mu$m contours (magenta), two 0.45\degr circles and a 0.9\degr circle (green). The locations of P1, P2, C1, C2, are marked by green crosses and magenta crosses separately. }
\label{fig:ROI2}
\end{figure}

Following the above works, we also tested several bubble-like templates (in addition to the disk emission, see \ref{fig:ROI2}): (1) two 0.45$\degr$ circles templates; (2) a single 0.45$\degr$ circle template in either side of the M31 disk; (3) a 0.9$\degr$ circle centered at M31; (4) two point sources, namely C1 and C2, located at the central positions of the 0.45$\degr$ circles. The spatial models are also displayed in the right panel of Figure~\ref{fig:ROI2}, and the results are listed in Table~\ref{tab:bubble}. We find that when using the two 0.45$\degr$ circle model and 0.9$\degr$ model, the TS values and the flux of the M31 disk are too small to be significant, which is unphysical, although these models {deliver higher} log$~\mathcal{L}$ than the single M31 models do (Table~\ref{tab:pass8}).  When using single bubble model (i.e., single 0.45$\degr$ circle template in either side of the M31 disk), the bubble templates have TS value smaller than 20.

Furthermore, we also tested an additional model including the M31 disk, the new source candidate P1, and P2 (as a hypothetical source located at an opposite side and same angular distance from the M31 disk as P1). In this case, P2 is not detected.

\section{Discussion}
We address the implications of our results in light of the possible diffuse components from M31: the disk, the bulge, and any bubble-like structure.

\subsection{Origin of $\gamma$-ray emission from the M31 bulge}

In Section 2.2.1, we have used a disk model, a bulge model, and a composed disk+bulge model as spatial models of M31, respectively.  Figure~\ref{fig:sed1} and Figure~\ref{fig:sed2} show the fitting results of the above three spatial models in the spectral energy distribution (SED) representation. 
In each fit, we divide the data into six logarithmic energy bins, covering the energy range $\rm 200 MeV-150 GeV$, as in \cite{yuan+etal+2014}. In Figure~\ref{fig:sed1}, we also plot the SEDs of M31 derived by \cite{abdo+etal+2010d} and \cite{li+etal+2016}, {both of which adopted a IRAS 100$\mu$m disk model.} The SED of the disk model agrees with their SEDs.

In the disk+bulge model, {assuming that} the $\gamma$-ray emission of the disk component originates from $\pi^0$ decay, and an injected proton spectrum of $dN/dE$ = $ \rm N_0$$\rm (1 + E/1.6GeV)^{2.8}$, which is {based on} the proton spectrum of the Milky Way, we plot the resulted $\gamma$-ray emission in both figures. We further assume that the bulge component of our disk+bulge model originates from MSPs, and plot a power law with exponential cutoff (PLE) spectrum with the parameters $\Gamma$ = 1.57, Ecut = 1.5 GeV, which is consistent with the typical MSPs in the Milky Way~(\citealp{abdo+etal+2009}). 
In Figure~\ref{fig:sed2}, we again overlay a hadronic spectrum for the disk model and a PLE spectrum (with $\Gamma$ = 1.0, $\rm E_{cut}$ = 1.1 GeV) for the bulge model.  

As referred in \ref{subsec:psf3}, we have detected a bulge component in disk+bulge model, with TS = 25.7, which we suggest as a strong evidence for the presence of the bulge component. We analogy this bulge $\gamma$-ray emission component to the Galactic Center Excess.

We examine the MSPs inside the bugle as the astrophysical origin of the bulge emission. Because M31 and the MW are local group galaxies that are comparable to each other, we can derive the number of MSPs needed for explaining the observed $\gamma$-ray luminosity of the bulge component, based on the luminosity function of Milky Way MSPs (\citealp{yuan+etal+2014}, \citealp{cholis+etal+2014}), 

\begin{equation}\label{equ:dndl}
dN/dL = kL^{-\alpha_1}[1+(L/L_{br})^2]^{(\alpha_1-\alpha_2)/2},
\end{equation}
where $\alpha_1$ = 1.1, $\alpha_2$ = 3.0,  $\rm L_{br}$ = 4 $\times 10^{33}\ {\rm erg\ s^{-1} }$, and k is the normalization factor. We thus have

 \begin{equation}\label{equ:ntot}
N_{tot}= \int_{L_1}^{L_2} kL^{-\alpha_1}[1+(L/L_{br})^2]^{(\alpha_1-\alpha_2)/2} dL,
\end{equation}
and
 \begin{equation}\label{equ:ltot}
L_{tot} = \int_{L_1}^{L_2} kL^{1-\alpha_1}[1+(L/L_{br})^2]^{(\alpha_1-\alpha_2)/2} dL,
\end{equation}
where $ \rm L_1$ = $10^{31}\ {\rm erg\ s^{-1} }$, $\rm L_2 $ = $10^{35}\ {\rm erg\ s^{-1} }$. $\rm L_{tot}$ is the $\gamma$-ray luminosity of M31 bulge, $\rm L_{tot}$ = (1.16$\pm$0.14)$\times10^{38}{\rm~erg~s^{-1}}$.  So, the number of the MSPs needed to produce the bulge component emission is $\rm N_{tot}\sim1.5\times10^5$. 

A more fundamental quantity is the MSP abundance, which is the number of MSPs divided by the underlying stellar mass. 
The stellar mass in the M31 bulge is estimated to be (2.5--6.6)$\times$10$^{10}{\rm~M_{\odot}}$ (Widrow et al. 2003, Tamm et al. 2012), resulting in a MSP abundance of (2--6)$\times$10$^{-6}$. For the MW bulge, the stellar mass is (0.5--2.7)$\times$10$^{10}{\rm M_{\odot}}$ (Licquia \& Newman 2015). 
Taking the ratio of the \gr~luminosity of the MW GeV excess ($\sim$2$\times$10$^{37}$erg~s$^{-1}$ from Bartels et al. 2017) and that of the M31 bulge derived in this work, which is about one-sixth, the number of MSPs in the MW (boxy) bulge is thus $\sim$2.5$\times$10$^4$, close to the value of (1--2)$\times$10$^4$ derived by Yuan \& Zhang (2014). \cite{eckner+etal+2018} estimate $\gamma$-ray emission of MSP population to explain the Galactic Center Excess and signal from center of M31, their conclusions also support our results.
This implies a MSP abundance of (1--5)$\times$10$^{-6}$ for the MW bulge. {\it Therefore, the MSP abundance is very similar between the bulges of M31 and MW, i.e., the ratio is close to unity.} We note that the ratio estimated here does not rely on the uncertainty of the luminosity function of MSPs, since it affects both estimated numbers of MSPs in the same manner. The major assumption here is that the \gr~emission predominantly arises from the MSP population in the bulges of both galaxies.

\subsection{Relation of $\gamma$-ray luminosity and IR luminosity of nearby galaxies}
\begin{figure}
\includegraphics[width=\textwidth]{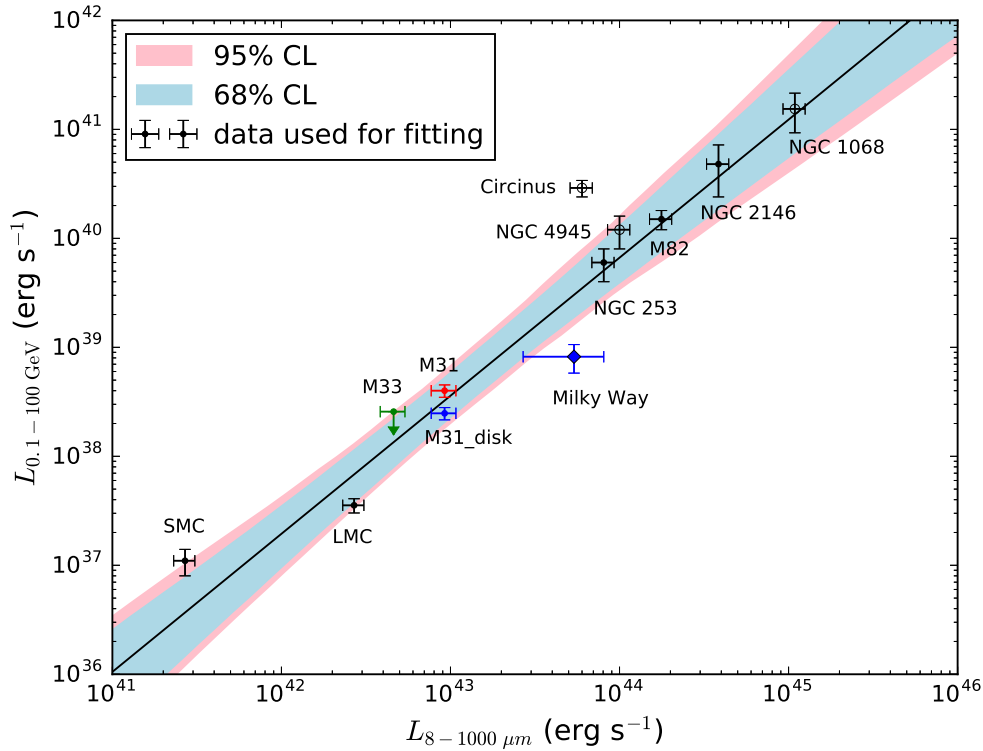}
\caption{Relation between the $\gamma$-ray luminosity ($\rm 0.1-100$ GeV) and total IR luminosity (8-1000~$\mu$m) for star-forming galaxies. {For LMC, the modified $\gamma$-ray luminosity is adopted, with the contributions of two bright pulsars subtracted from total luminosity. The black solid line represents the best-fit result IR-$\gamma$ relation including NGC\,253, NGC\,2146, M82, SMC, LMC, with a slope of 1.21 $\pm$ 0.11. The pink/blue shade represents the fitting results at 95$\%$/68$\%$ confidence.}}
\label{fig:gi}
\end{figure}

\begin{figure}[tbh!]
\centerline{
\includegraphics[width=0.5\textwidth]{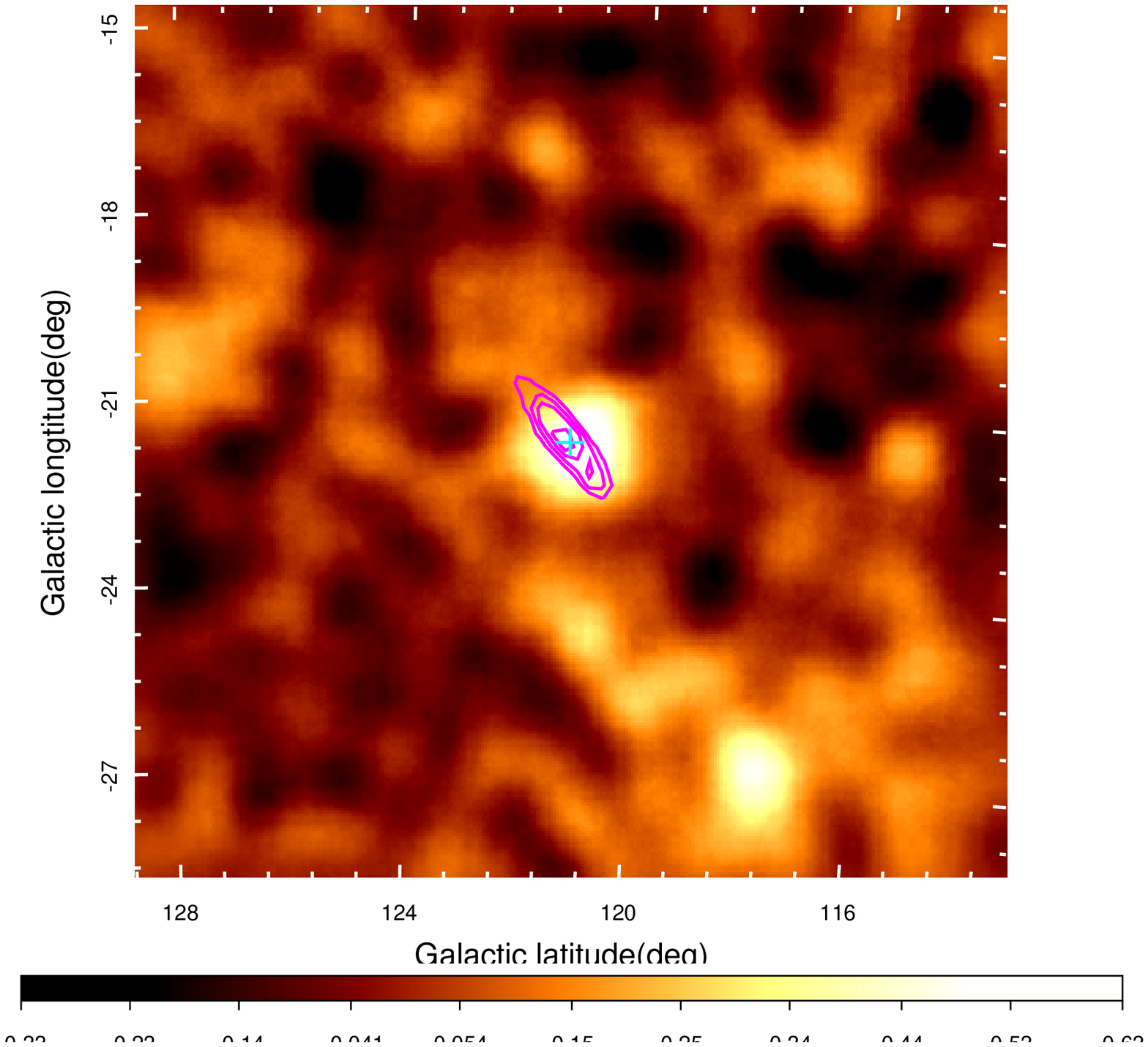}
\includegraphics[width=0.5\textwidth]{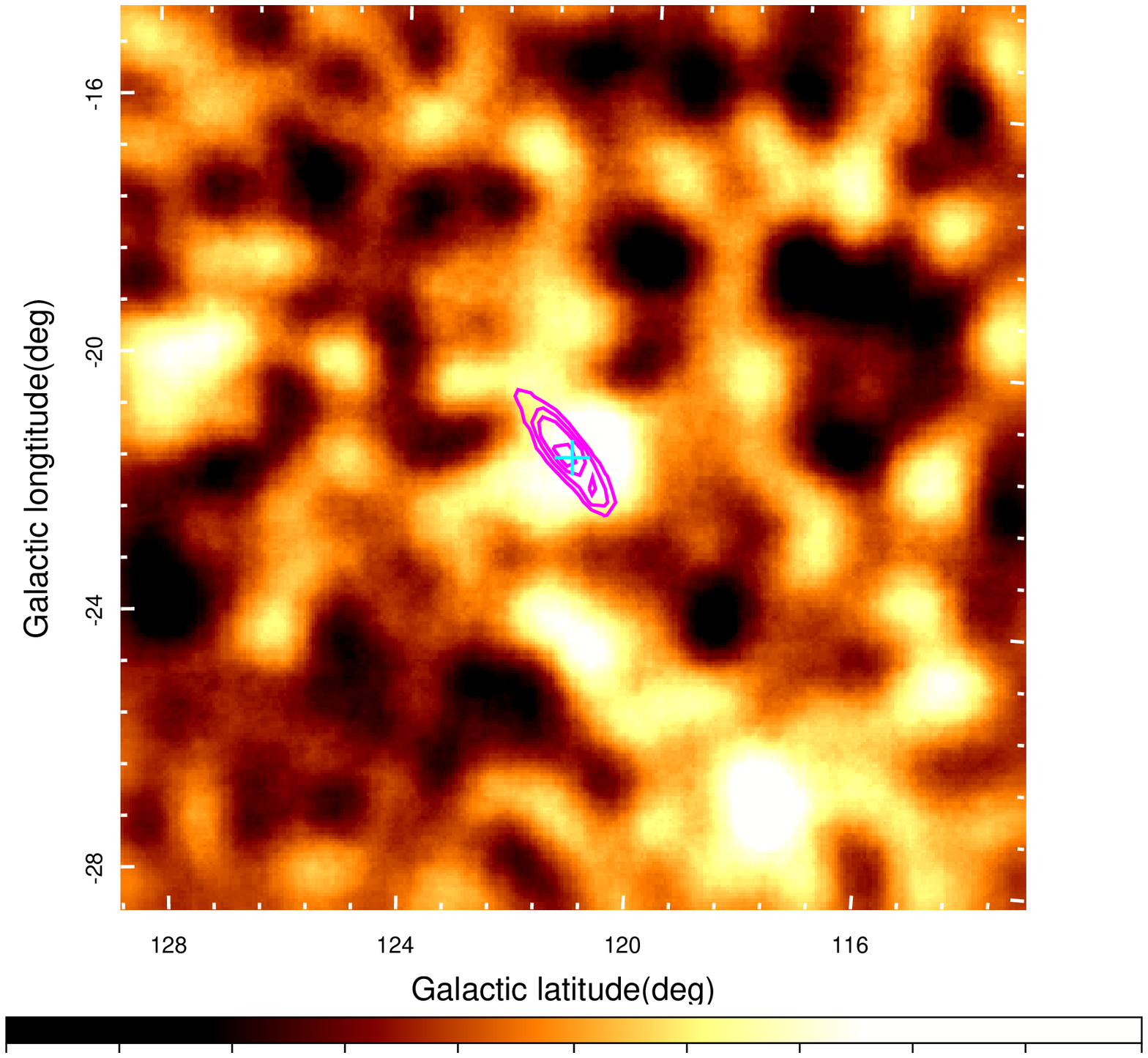}
}
\vskip0.7mm
\centerline{
\includegraphics[width=0.5\textwidth]{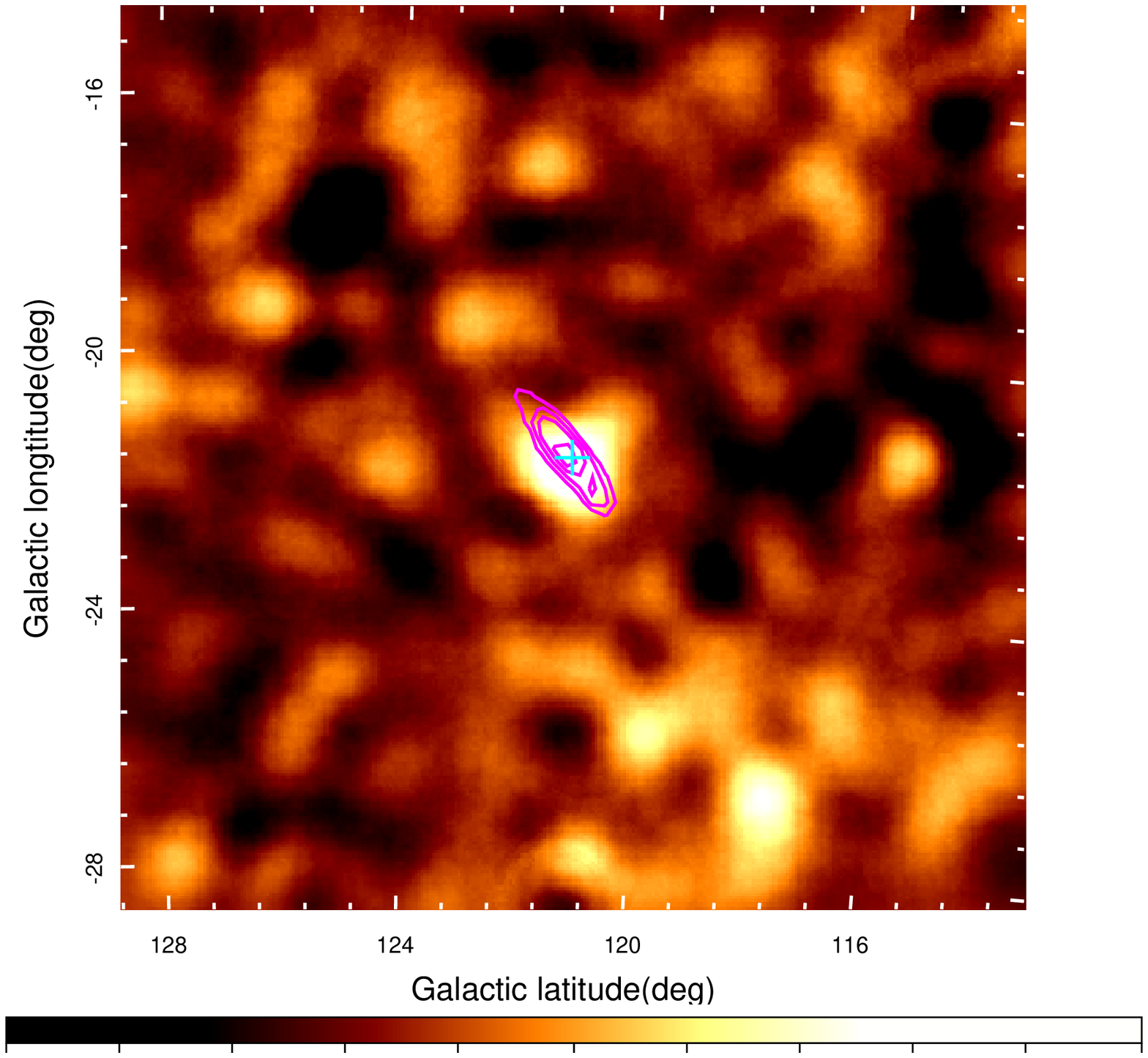}
\includegraphics[width=0.5\textwidth]{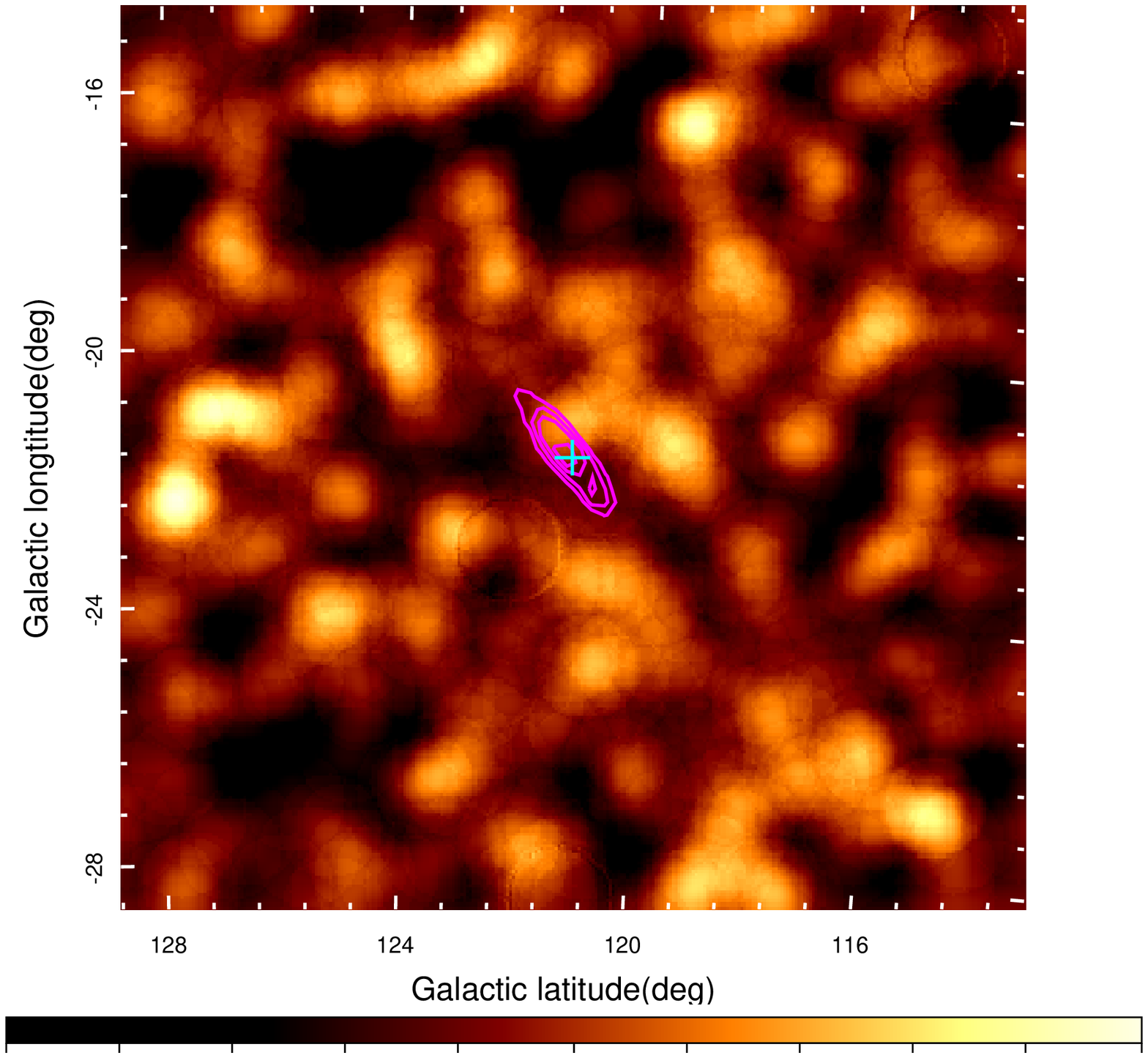}
}
\caption{Residual counts maps of different energy sub-bands.
{\it Top left:} 200 MeV--300 GeV; {\it Top right:} 200 MeV--1 GeV; {\it Bottom left:} 1--20 GeV; {\it Bottom right:} 20--300 GeV.
In all panels, background sources have been subtracted.
Prominent emission is seen coincident with M31 except in the 20--300 GeV band. All maps are smoothed with a Gaussian kernel of 0.8\degr~and overlaid with the IRAS 100~$\mu$m intensity contours (magenta). The cyan cross marks the center of M31.}
\label{fig:subband}
\end{figure}

\begin{figure}
\includegraphics[width=\textwidth]{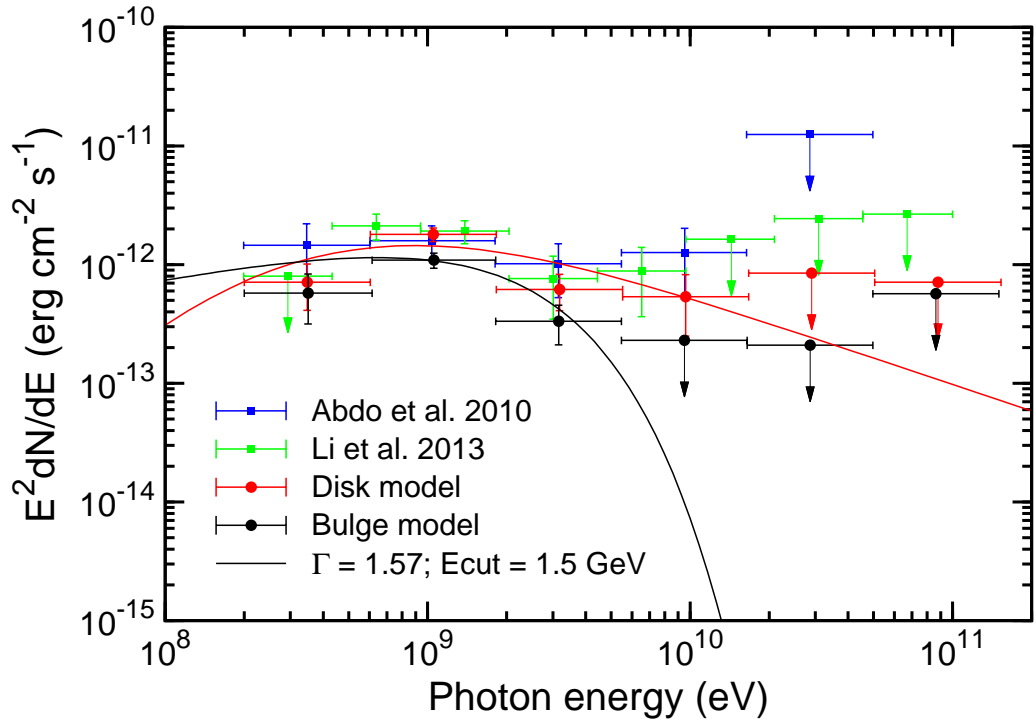}
\caption{SEDs of the disk (red points) and bulge (black points) models. The red solid curve represents the $\gamma$-ray spectrum expected from a proton spectrum of the form $dN/dE$ = $ \rm N_0$$\rm (1 + E/1.6GeV)^{2.8}$, where E is the kinetic energy of the protons, while the black solid curve represents a PLE spectrum characteristic of MSP spectra. The SEDs derived by Abdo et al. 2010 (blue) and Li et al. 2016 (green) are plotted for comparison.}
\label{fig:sed1}
\end{figure}

\begin{figure}
\includegraphics[width=\textwidth]{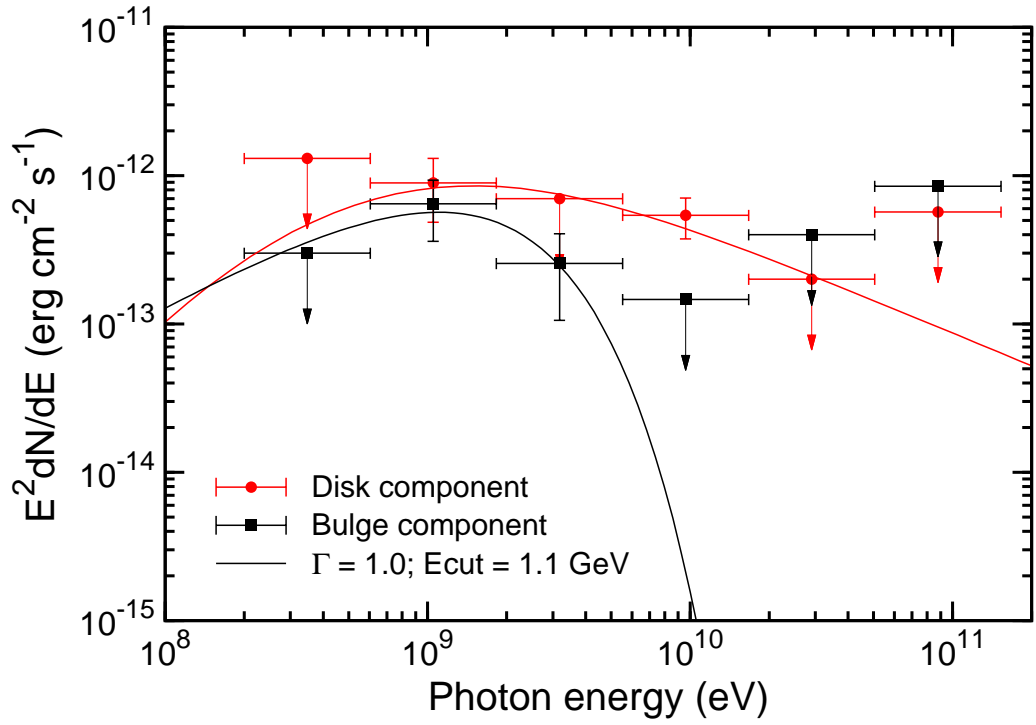}
\caption{SED of the disk+bulge model. The SEDs of the disk component (red points) and bulge component (black points) are displayed separately. The red solid curve represents the $\gamma$-ray spectrum expected from a proton spectrum of the form $dN/dE$ = $ \rm N_0$$\rm (1 + E/1.6GeV)^{2.8}$, where E is the kinetic energy of the protons, while the black solid curve represents a PLE spectrum characteristic of MSP spectra.}
\label{fig:sed2}
\end{figure}

Figure~\ref{fig:gi} shows the correlation between the $\gamma$-ray luminosities and the total infrared (IR) luminosities (8-1000 $\mu$m) of several nearby galaxies, including local group galaxies, star forming galaxies and AGNs (\citealp{abdo+etal+2010a}, \citealp{tang+etal+2014}). {However, here we} fit this IR-$\gamma$ correlation using {only} star forming galaxies SMC, LMC, NGC 253, M~82, NGC 2146. The best fit is plotted as the black solid line in Figure~\ref{fig:gi}, with a slope of 1.21$\pm$0.11. We exclude NGC 1068, NGC 4945, and Circinus galaxy in the fitting, as their $\gamma$-ray emissions are probably dominated by AGNs. We also exclude Milky Way from the fit, due to the possible underestimation of its total $\gamma$-ray luminosity because of our internal perspective. The total 8-1000 $\mu$m luminosities of the Circinus galaxy is taken from \cite{ha+etal+2011}, those of other galaxies are taken from \cite{gao+etal+2004}. The $\gamma$-ray luminosity of $\rm NGC~2146$ is taken from \cite{tang+etal+2014}. The $\gamma$-ray luminosities of other galaxies are taken from \citealp{ack+etal+2012a}. Two remarkable PSRs have been recognized in LMC recently (Ackermann et al. 2016). PSR J0540-6919 has a $\rm L_{0.1-100 GeV} = (5.88\pm1.36)\times 10^{36}\ {\rm erg\ s^{-1}}$ and PSR J0537-6910 has  a $\rm L_{0.1-100 GeV} = (5.6\pm1.02)\times 10^{36}\ {\rm erg\ s^{-1} }$. Their total luminosity L = (1.15$\pm$0.17)$\times 10^{37}\ {\rm erg\ s^{-1} }$, accounts for about 24\% of the total $\gamma$-ray luminosity of LMC. We subtract the $\gamma$-ray contribution from these two PSRs. This modification of LMC luminosity has tiny influence on the fit of the IR-$\gamma$ relation of galaxies.

\begin{table*}
\begin{minipage}[]{160mm}
\begin{center}
\caption{Summary of sample galaxies}
\label{tab:galaxies}
\begin{tabular}{cccccccccccc}
\hline\hline
Galaxy &
distance &
Index &
 TS &
{$L_{\rm 0.1-100~GeV}$} &
$L_{\rm 8-1000~{\mu}m}$
\\
(1) &
(2) &
(3) &
(4) &
(5) &
(6) \\
\hline\hline

SMC                  & 0.06  &  2.22$\pm$0.02   & 136.6  &  0.11$\pm$ 0.03   &0.007$\pm$0.001  \\
LMC                  & 0.05  &  2.02$\pm$0.02   &1122    & 0.47$\pm$0.05     & 0.07$\pm$0.01  \\
LMC*                  & 0.05  &  2.02$\pm$0.02   &1122    & 0.35$\pm$0.05    & 0.07$\pm$0.01  \\
M33                  &0.85   &  2.48$\pm$0.06  & 13.65  & $\lesssim$3.5     &  0.12$\pm$0.02   \\
M31                   & 0.78  & 2.31$\pm$0.09    & 107.87  &  5.15$\pm$0.5    & 0.24$\pm$ 0.04 \\
$\rm M31^{a}$        & 0.78  &  2.22$\pm$ 0.13 & 33.4  & 3.84$\pm$0.03    & 0.24$\pm$ 0.04 \\
Milky Way             &  --  & 2.2$\pm$0.1      &  --    & 8.2$\pm$2.4      & 1.4$\pm$ 0.7\\
NGC 253               & 2.5  & 2.2$\pm$0.1      & 109.4  & 60$\pm$20        & 2.1$\pm$0.32  \\
NGC 4945              & 3.7  & 2.1$\pm$0.2      & 33.2   & 120$\pm$40       & 2.6$\pm$0.39\\
M82                   & 3.4  &  2.2$\pm$0.1     & 180.1  & 150$\pm$30       & 4.6$\pm$0.69\\
NGC 2146              & 15.2 & 2.2$\pm$0.1      &  30.8  & 400$\pm$210      & 10$\pm$1.5 \\
NGC 1068              &16.7  & 2.2$\pm$0.2      &38.1    & 1540$\pm$610     & 28.3$\pm$ 4.25\\
Circinus               &4.2  & 2.19$\pm$0.12    & 58     &290$\pm$50        & 1.56$\pm$ 0.23 \\
\hline
\end{tabular}
\end{center}
(1) Galaxy name; (2) Distance, in units of Mpc; (3) Photon-index of $\gamma$-ray emission; (4) TS values; (5) {$0.1-100$ GeV luminosity}, in units of $10^{38}{\rm~erg~s^{-1}}$; (6) Total IR (8--1000~$\mu$m) luminosity, in units of $10^{10}~L_\odot$. {``M31" refers to M31 disk model, while ``$\rm M31_{disk}$''} refers to the disk component of disk+bulge model. The Circinus galaxy is taken from \cite{ha+etal+2011}. For other galaxies, the total IR luminosities are taken from \cite{gao+etal+2004} and the $\gamma$-ray luminosities are taken from \cite{ack+etal+2012a}. The $\gamma$-ray luminosity of NGC\,2146 is from \cite{tang+etal+2014}.
\end{minipage}
\end{table*}

From Figure 8, we find that most of the galaxies in our sample lie within the 95$\%$ confidence level of the best fit line. This is consistent with \cite{abdo+etal+2010d}. According to Figure~\ref{fig:gi}, the $\gamma$-ray luminosity of the disk (disk component) agree with the IR-$\gamma$ correlation well in {both the disk and the} disk+bulge model. The relationship of $\rm L_{IR}$ and $\rm L_{\gamma}$ holds for different galaxies including star forming galaxies and star burst galaxies, which may indicate the {dominating} effects of acceleration of protons by star forming regions, and/or related to the nature of CR electron calorimetry~(\citealp{mur+etal+2006}).

\subsection{Non-detection of bubble-like features of M31} \label{subsec:bubble}

We tested several bubble-like templates as mentioned in previous works \cite{psh+etal+2016} and  \cite{li+etal+2016}. We did not find any significant emission like those claimed in \cite{psh+etal+2016} and we confirm the non-detection of such structures as in \cite{li+etal+2016}.

As mentioned in Section~\ref{subsec:psf3}, we found a source candidate P1 $\sim$ 40 arcmin northwest to M31. We then add P1 into the {spatial models}. {A point source and a power-law are used as the spatial model and the spectral model of P1, respectively}. {We redo the likelihood analysis with the full pass8 data (Table~\ref{tab:pass8}) and the PSF3 data (Table~\ref{tab:psf3}), separately.} The TS values of P1 {in the analysis with the full data} are just below the detection threshold 25. We suggest it is a new source candidate.  After P1 is added to the {spatial models}, the log$~\mathcal{L}$ values of the models are improved. P1 could be a background source, or a source connected to M31. If the latter one is the case, it will be very interesting. This emission could be related to the past activities of the nucleus of M31. To test this assumption, we place a hypothetical source located at an opposite side and same angular distance of the M31 disk as P1 (which we call P2). P2 is not detected in our analysis, thus P1 lacks a symmetric geometrical counterpart which could strengthen P1 as a Fermi bubble-like feature.

The two Fermi bubbles of the Milky Way have a luminosity $L_{\rm 1-100GeV} = 4\times10^{37}{\rm erg\ s^{-1} }$ with a spectral index $\sim$ 2~\citep{su+etal+2010}, which is about 5\% of the total Galactic $\gamma$-ray luminosity of 0.1-100 GeV~\citep{strong+etal+2010}. We obtain the luminosity of the residual point source P1 in the same energy range $L_{\rm 1-100 GeV}$ = (5.4 $\pm$ 2.1)$\times 10^{37}\ {\rm erg\ s^{-1}}$, assuming P1 is associated with M31 at a distance 780 kpc. The total $\gamma$-ray luminosity of M31 is (4.0$\pm$ 0.5)$\times 10^{38}\ {\rm erg\ s^{-1} }$ (for disc model), and the P1 source is about 5 -- 21\% of the total luminosity. P1 is located at a distance $\sim$10.5 kpc to the M31 disk, which is comparable to the distance $\sim$4 kpc from the center of the Fermi bubbles to the Galactic disk. We plot the spectrum of P1 in Figure \ref{fig:ROI2}. The spectral index of P1 $\sim$ 2.31 is slightly softer than {the one of the Fermi bubble} by Su et al (2010).

\section{Summary}

In this work we study the $\gamma$-ray emission of M31 by using more than 8 years of {Fermi-LAT data, which includes the full pass8 data and the PSF3 only data in the energy range from 200~MeV to 300~GeV. We have used the disk model, bulge model and disk+bulge model to represent the spatial distribution of M31, where the disk component in these models are based on IRAS 100~$\mu$m image.} Our findings are summarized below:

\begin{enumerate}
\item We find that disk+bulge model provides the highest log$~\mathcal{L}$, that means it is the best among the tested models. In this case $\rm TS_{ disk}$ = 33.4, while $\rm TS_{ bulge}$ = 25.7, {suggesting} a strong evidence for the detection of both the central bulge component and the disk component of M31. 
\item Assuming a major fraction of the bulge-like \gr~emission is originated from MSPs, we calculate the number of MSPs needed to explain the luminosity of the bulge component {in the disk+bulge model} is $\rm N_{tot}\sim1.5\times10^5$. The thus derived MSP abundance of the M31 bulge, $(2-6)\times10^{-6}$, is close to the value of $(1-5)\times10^{-6}$ for the MW bulge, provided that the $\gamma$-ray luminosity of the latter is also dominated by MSPs. 

\item M31 disk model and M31 disk component of the disk+bulge model both satisfy the relation between the $\gamma$-ray luminosity ($\rm 0.1-100$ GeV) and total IR luminosity (8-1000~$\mu$m) for star-forming galaxies.

\item We analysed pass8 PSF3 data, which are preselected data sets with the best angular resolution. We found a source candidate P1 located about $0.7\degr $ northwest to M31, with a significance $\sim$ 4.7~$\sigma$ in the full-data set analysis. There is no source coincident with P1 in FL8Y catalogue.
\item We did not find any significant bubble-like features in the region of M31. If we compare P1 to Fermi bubble, the total luminosity of P1 in $1-100$ GeV is similar to that of the Fermi bubble. Both of them have {shown similar fractions of the total luminosities of their host galaxies, which is $\sim5\%$. We didn't find counterpart of P1 on the southeast side of M31. }
\end{enumerate}

\section*{Acknowledgments}
We thank Xian Hou, Yudong Cui and Xiang-Dong Li for valuable comments. This work uses data and software provided by Fermi Science Support Center. This work was supported by the 973 Program under grants 2017YFA0402600 and 2015CB857100 and the National Natural Science Foundation of China under grants 11473010, 11133001, 11773014, 11633007 and 11851305.


\begin{table*}
\begin{center}
\caption{Centroid of M31}
\begin{tabular}{ccccc}
\hline\hline
Energy Band & RA & DEC & Error \\
 & (degree)  & (degree) & (degree) \\
(1) & (2) & (3) & (4) & \\
  \hline
  \hline
0.2-300 GeV (this work)    &  10.7806   &  41.2742  & 0.0894   \\

\hline
1-300 GeV (this work)     & 10.8466  &   41.2223  & 0.0769  \\
\hline
1-100 GeV        & 10.81  & 41.19 & 0.07 (in RA)  \\
(\citealp{ack+etal+2017}) & & & 0.05 (in DEC)\\
\hline
optical   & 10.6847 & 41.2687 \\
\hline
\label{tab:test}
\end{tabular}
\end{center}
\end{table*}

\begin{table*}
\caption{Likelihood analysis for sub-bands}
\begin{tabular}{cccccccccccc}
\hline
\hline
Energy band  & spatial model & composition & $\Gamma$ & $\rm F_{0.2-300~GeV}$  & $\rm L$  & TS & log$~\mathcal{L}$ \\
 & & &  & ($10^{-9}{\rm~ph~s^{-1}~cm^{-2}}$) & ($10^{38}{\rm~erg~s^{-1}}$)  & & &  \\
(1) & (2) & (3) & (4) & (5) & (6) & (7) &  (8)  \\
\hline
\hline
0.2-1 GeV  & disk   & IRAS 100$\mu$m     & 1.39$\pm$0.34    & 2.5$\pm$0.50      &  1.32$\pm$ 0.22    &41.44   &  -614007.8     \\
                 &bulge  &  & 1.70$\pm$0.44    &  1.81$\pm$ 0.50   &   0.89$\pm$0.12   & 28.0   &  -614008.9    \\
             &disk+bulge  & IRAS 100$\mu$m   & 1.31$\pm$ 0.56   &  1.71$\pm$0.56   &   0.92$\pm$0.24   & 19.86    &   -614007.6\\
               &       &  bulge      & 1.85$\pm$ 0.97     & 0.69$\pm$0.46   &   0.33$\pm$0.19 &  3.41   &   -      \\
\hline
1-20 GeV       & disk  & IRAS 100$\mu$m  & 2.75$\pm$0.16      & 0.60$\pm$0.09  & 1.31$\pm$0.20   &53.45 &   -180236  \\
                 &bulge  &  &  3.42$\pm$0.18    &  0.36$\pm$0.06 &  0.7$\pm$0.14  & 52.08 &   -180233       \\
                & disk+bulge& IRAS 100$\mu$m      & 2.15$\pm$ 0.33   &  0.24$\pm$0.08   &   0.81$\pm$0.30    & 9.56    &    -180231\\
                 &    & bulge        & 3.66$\pm$ 0.63     & 0.27$\pm$0.06   &   0.49$\pm$0.14   &  26.25  &        -      \\
\hline
\label{tab:subband}
\end{tabular}
\end{table*}

\begin{table*}
\caption{Likelihood analysis for bubble-like templates}
\begin{tabular}{cccccccccccc}
\hline\hline
spatial model &
composition &
$\Gamma$  &
$\rm F_{0.2-300~GeV}$ &
$\rm L$  &
TS value  &
log$~\mathcal{L}$ \\
 &
 &
 &
($10^{-9}{\rm~ph~s^{-1}~cm^{-2}}$) &
($10^{38}{\rm~erg~s^{-1}}$) &
 &
\\
(1) &
(2) &
(3) &
(4) &
(5) &
(6) &
(7) \\
\hline\hline
 disk+two 0.45\degr circles   &      IRAS 100$\mu$m       & 2.16$\pm$0.22    & 1.34$\pm$1.06   &  1.53$\pm$0.93 &  11.64  &  -501376   \\
                    & two 0.45\degr circles  &         2.29$\pm$0.10   & 3.53$\pm$0.90 &  3.19$\pm$0.77    & 64.0   &-\\
 \hline
disk+0.45\degr circle1    &     IRAS 100$\mu$m &2.25$\pm$0.12 & 3.05$\pm$0.83   & 2.96$\pm$0.66  & 52.0    & -501382    \\
                     & 0.45\degr circle1   &  2.34$\pm$0.19 &   1.67$\pm$0.68 &   1.40$\pm$0.46   &   18.54   &- \\
\hline
disk+0.45\degr circle2  &   IRAS 100$\mu$m        & 2.3$\pm$0.12 & 3.48$\pm$0.55   & 3.06$\pm$0.51    &  60.42    &  -501383 \\
                     & 0.45\degr circle2  &2.19$\pm$0.18 &  1.17$\pm$0.41 &  1.27$\pm$0.44 & 12.53 &- \\
\hline
disk+0.9\degr circle      &   IRAS 100$\mu$m    &  2.22$\pm$0.27    &  0.84$\pm$0.49   & 0.86$\pm$0.49   &   3.95&  -501378    \\
                         &0.9\degr circle  &   2.21$\pm$0.09    &  4.35$\pm$0.56   &  4.51$\pm$0.68    & 87.48 &   -     &\\
\hline
disk+P1+P2 & IRAS 100$\mu$m &  2.25$\pm$0.12   &    3.66$\pm$0.4   &  3.53$\pm$0.56        &    76.87     &  -501376         \\
            & P1& 2.32$\pm$0.18         &     1.14$\pm$0.26 &  0.97$\pm$0.3 &       23.52     &  -         \\
            & P2&  2.33$\pm$0.17  &     0.3$\pm$0.02   &  0.25$\pm$0.03     &     1.46      &   -        \\

\hline
disk+C1+C2 & IRAS 100$\mu$m &   2.2$\pm$0.13           &     2.85$\pm$0.37        &  3.05$\pm$0.57        &    51.18   &       -501382     \\
            & C1 & 2.70$\pm$0.15  &   1.34$\pm$0.4      & 0.74$\pm$0.2        &  10.24 &    -   \\
            & C2 &  2.35$\pm$0.11      &      0.64$\pm$0.03   &  0.52$\pm$0.03     &   5.32    &  - \\

\hline
\label{tab:bubble}
\end{tabular}
\end{table*}

\end{document}